\documentclass[%
 aip, amsmath,amssymb,
 reprint,%
]{revtex4-1}

\usepackage{graphicx}
\usepackage{dcolumn}
\usepackage{bm}

\usepackage[utf8]{inputenc}
\usepackage[T1]{fontenc}
\usepackage{mathptmx}
\usepackage{verbatim}
\usepackage{fdsymbol}
\usepackage{graphicx}
\usepackage{sidecap}
\usepackage{placeins}
\usepackage{xcolor}

\makeatletter
\DeclareTextCommand{\textprime}{\encodingdefault}{%
  \mbox{$\m@th'\kern-\scriptspace$}%
}
\makeatother

\begin{document}

\preprint{AIP/123-QED}

\title[Dynamics and bifurcations in multistable 3-cell neural networks]{Dynamics and bifurcations in multistable 3-cell neural networks}

\author{J. Collens}
\affiliation{ Neuroscience Institute, Georgia State University, Atlanta, GA, 30303, USA}
\affiliation{Department of Mathematics and Statistics, Georgia State University, Atlanta, GA, 30303, USA}
\author{K. Pusuluri}
\affiliation{ Neuroscience Institute, Georgia State University, Atlanta, GA, 30303, USA}
\affiliation{Corresponding author, email: pusuluri.krishna@gmail.com}
\author{A. Kelly}
\author{D. Knapper}
\affiliation{ Neuroscience Institute, Georgia State University, Atlanta, GA, 30303, USA}
\author{T. Xing}
\affiliation{Department of Mathematics and Statistics, Georgia State University, Atlanta, GA, 30303, USA}
\author{S. Basodi}
\affiliation{Department of Computer Science, Georgia State University, Atlanta, GA, 30303, USA}
\author{D. Alacam}
\affiliation{Department of Mathematics and Statistics, Georgia State University, Atlanta, GA, 30303, USA}
\affiliation{Department of Mathematics, Bursa Uludağ University, Bursa, 16059, Turkey}
\author{A. Shilnikov}
\affiliation{ Neuroscience Institute, Georgia State University, Atlanta, GA, 30303, USA}
\affiliation{Department of Mathematics and Statistics, Georgia State University, Atlanta, GA, 30303, USA}

\date{\today}

\begin{abstract}

We disclose the generality of the intrinsic mechanisms underlying multistability in reciprocally inhibitory 3-cell circuits composed of simplified, low-dimensional models of oscillatory neurons, as opposed to those of a detailed Hodgkin-Huxley type \citep{wojcik2014key}. The computational reduction to return maps for the phase-lags between neurons reveals a rich multiplicity of rhythmic patterns in such circuits. We perform a detailed bifurcation analysis to show how such rhythms can emerge, disappear, and gain or lose stability, as the parameters of the individual cells and the synapses are varied.

\end{abstract}

\maketitle

\textbf{Complex multistable bursting rhythms can arise even in simple biological neural circuits. We employ a computational technique combining phase-lag return maps and fast (parallel) GPU-based sweeps of the phase and parameter spaces, to identify multistable patterns, rhythm switching, and attractor robustness in reciprocally inhibitory 3-cell circuits composed of the proposed generalized Fitzhugh-Nagumo model of ``spike-less” bursters. With such maps, we can thoroughly examine how internal and external factors such as the synaptic strengths, network asymmetry, and externally injected currents determine what stable rhythmic patterns can co-exist, emerge or disappear, as well as their underlying bifurcation mechanisms. Depending on intrinsic mechanisms, such as release and escape in individual cells, these networks can produce a variety of multistable rhythmic states, ranging from penta-stability with phase-locked bursting pacemakers and traveling wave patterns, or stable chimeras admitting recurrent phase slipping of one cell with respect to the other two that remain phase-locked over time, to more exotic behaviors such as a robust stable synchronous state with all three cells oscillating together, or a lack of phase locked rhythmic states altogether. We present detailed transition mechanisms between such rhythms including saddle-node, pitchfork, and secondary Andronov-Hopf or torus bifurcations, as well as the emergence of a transitive torus. Lastly, we introduce the concept of $2\theta$-neurons to build even simpler neural circuits capable of desired dynamics. Our qualification promotes the use of simplified, low-dimensional modeling of multistable bursting patterns arising in oscillatory neural circuits, in lieu of computationally intensive high-dimensional Hodgkin-Huxley type models. }

\section{\label{sec:Introduction} Introduction}

A central pattern generator (CPG) is a small network of coupled neurons that determines and autonomously controls rhythmic oscillations underlying sensory, motor or cognitive behaviors of an animal. CPGs are implicated in a variety of functions ranging from respiration, heartbeat and circulation, to sleep and locomotion \cite{miller1985neural,bal1988pyloric,marder1996principles, kristan2005neuronal,calin2007parameter,sherwood2011synaptic, newcomb2012homology,milo2002network,sporns2004motifs, rabinovich2006dynamical,bulloch1992reconstruction, marder1994invertebrate,frost1996single,katz2016evolution}. While many insights into the operational principles of CPGs have been obtained from experimental and computational studies, the basic principles of robustness and stability of many CPGs observed in nature are yet poorly understood and cannot be inferred {\em a priori.} The cooperative dynamics of coupled cells is an area of active ongoing research, with both biological and phenomenological approaches employed \cite{alaccam2015making,brown1911intrinsic,jalil2013toward, newcomb2012homology, sakurai2011distinct, sakurai2011different, sakurai2014two, katz1998comparison, sakurai2017artificial}. The smallest building unit commonly tied to many CPGs is a pair of bilaterally symmetric neurons that mutually inhibit each other to produce anti-phase bursting, called a {\em half-center oscillator} (HCO). The current study is focused on the rhythmic dynamics, transitions and bifurcations occurring in the context of a 3-cell neural network motif, made up of interconnected HCO circuits. Various 3-cell biological circuits that form constituent blocks or centers for larger networks have been reported previously. \cite{kopell2004chemical,matsuoka1987mechanisms,kopell1988toward, canavier1994multiple, skinner1994mechanisms,dror1999mathematical, prinz2003alternative}

\begin{figure*}[ht!]
\begin{center}
\resizebox*{1.6\columnwidth}{!}{\includegraphics{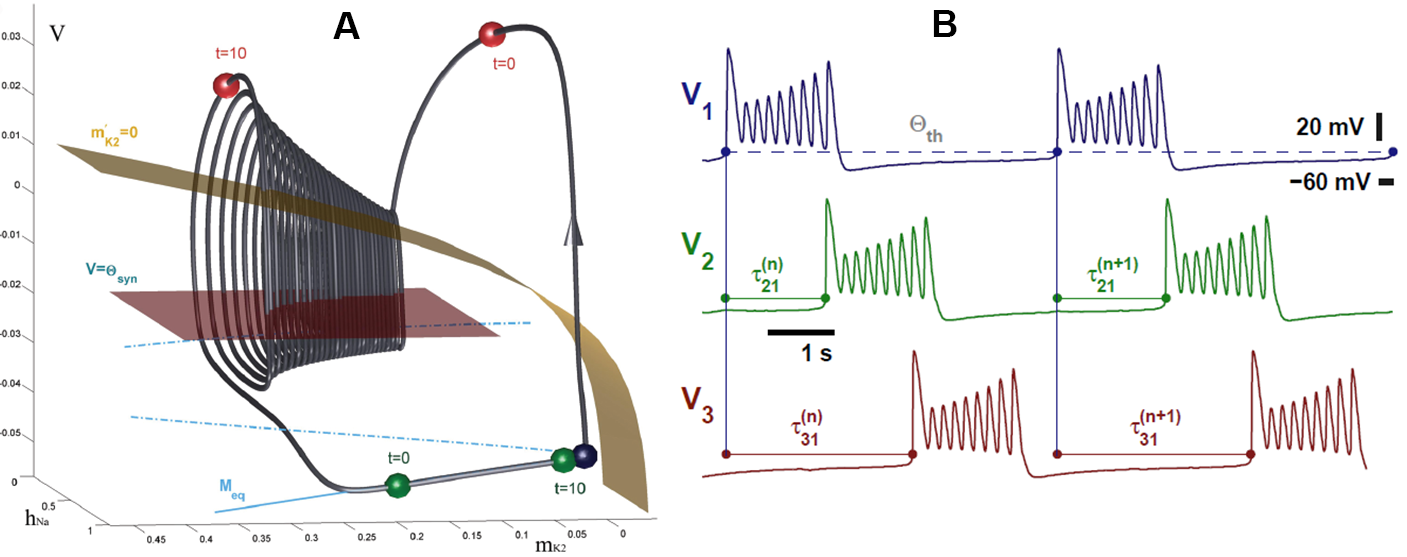}}
\caption{ 
   \textbf{(A)} Snapshot depicting the current states (represented by the blue, green and red spheres) of three weakly-coupled cells at $t=0$ and their further progressions at $t=10$, on the bursting orbit (grey) in the 3D phase space of the Hodgkin-Huxley type model of the leech heart interneuron \cite{rcd2004,shilnikov2012complete}. The plane $\Theta_{\rm syn}$ represents the threshold for the chemical synapses, that divides the active ``on'' phase (above it) and the inactive ``off'' phase; here, the active red cell inhibits the quiescent  green and blue ones. \textbf{(B)} Burst initiations in successive voltage traces generated by the 3-cell neural network allow us to define the relative delays $\tau$'s and hence, the phase lags (given by Eqs.~(\ref{phaselageq})) between its constituent bursters; see \cite{wojcik2011order,wojcik2014key} and the Methods section below, for further details. }
\label{fig:HH}
\end{center}
\end{figure*}

Several modeling paradigms have been applied for studies of such circuits, including biologically relevant Hodgkin-Huxley (HH) type models, in which the individual parameters can be related to specific ionic currents or concentration gradients. The high dimensionality of the detailed HH type models presents obvious difficulties in performing a thorough dynamical and bifurcation analysis to classify their generic properties. Such understanding is essential to reliably assemble and configure small neural networks with common oscillatory characteristics. The simpler integrate-and-fire models, which belong to the opposite end of the spectrum of mathematical models, are often inadequate to connect their parameters to biological mechanisms that may be directly manipulated or affected, failing to capture nuances in the dynamic behaviors that intermediate systems can. In this paper, we employ the so-called  {\em generalized} FitzHugh-Nagumo (gFN) neurons to model 3-cell networks. This 2D gFN-neuron model presents a better description of some of the pivotal properties of typical HH-type square wave bursters. We use it to showcase the essential characteristics of the building blocks of rhythmic circuits to stably generate the desired dynamics, regardless of the specific models of neurons and synapses. The gFN-equations are simpler and hence more practical for computational studies, especially for intense GPU-based sweeping of parameters and initial conditions.      

	This paper capitalizes on our previous work and the well established principles in the characterization of 3-cell circuits made of HH-type neurons (as depicted in Fig.~1) \cite{shilnikov2008polyrhythmic,wojcik2011order,wojcik2014key}. The goal is to present novel findings using parameter sweeps that reveal how the multistability of a 3-gFN motif is aligned with its parameter space. It also serves as a tutorial blueprint, providing a complete framework for interested researchers to borrow and employ our methods for the analysis of similar oscillatory networks. We employ gFN-neurons to study a variety of polyrhythmic dynamics arising in 3-cell circuits with symmetric and asymmetric connectivity, as well as various fast, slow and delayed effects. While maintaining generic behaviors, this reduction in complexity allows for more extensive exploration of key parameters in neural systems ranging from inherently quiescent or tonic spiking, to intrinsic bursters. It also aids our search for biologically plausible circuitries that ensure the robustness of the rhythmic patterns observed in nature. One important aspect of the so-called multifunctional CPGs is the ability of the same circuit to produce more than one observable rhythmic outcomes, as well as to switch between its rhythms \cite{dror1999mathematical,wojcik2011order,rubin2012explicit, calabrese2011coping, kristan2008neuronal,briggman2008multifunctional, schwabedal2016qualitative, xing2015computational}. We examine how changes in parameters can trigger transitions or bifurcations in the rhythms of an otherwise robust network. We summarize the use of phase lags, their corresponding Poincar\'e return maps, and fast GPU-based parallel simulations to characterize the state space of the 3-cell motif and to identify its stable and unstable rhythms. 

Specific results are presented in terms of two primary mechanisms underlying network rhythmogenesis -- synaptic release and escape \cite{wang1992alternating, schwabedal2016qualitative}. For each mechanism, we contrast the behaviors of symmetric networks with uniform all-to-all inhibitory connections, to those observed in asymmetric motifs with one or several distinct connections. We study how network circuitry influences the occurrence and stability of various rhythms with phase-locked states, periodic phase slipping or chimera-like behaviors. We determine the ranges of network parameter values, specifically the synaptic coupling strengths and the external current drives, that give rise to these behaviors and the underlying bifurcations. We also connect overarching mechanisms and present the stages of transition from the release to the escape mechanisms, in terms of bifurcation sequences that the corresponding fixed points (FPs) in the Poincar\'e return maps for the phase lags between the constituent neurons may undergo due to network parameter variations. We elaborate on several specific examples chosen for both relevance and novelty.

\begin{figure*}[ht!]
\begin{center}
\resizebox*{1.7\columnwidth}{!}{\includegraphics{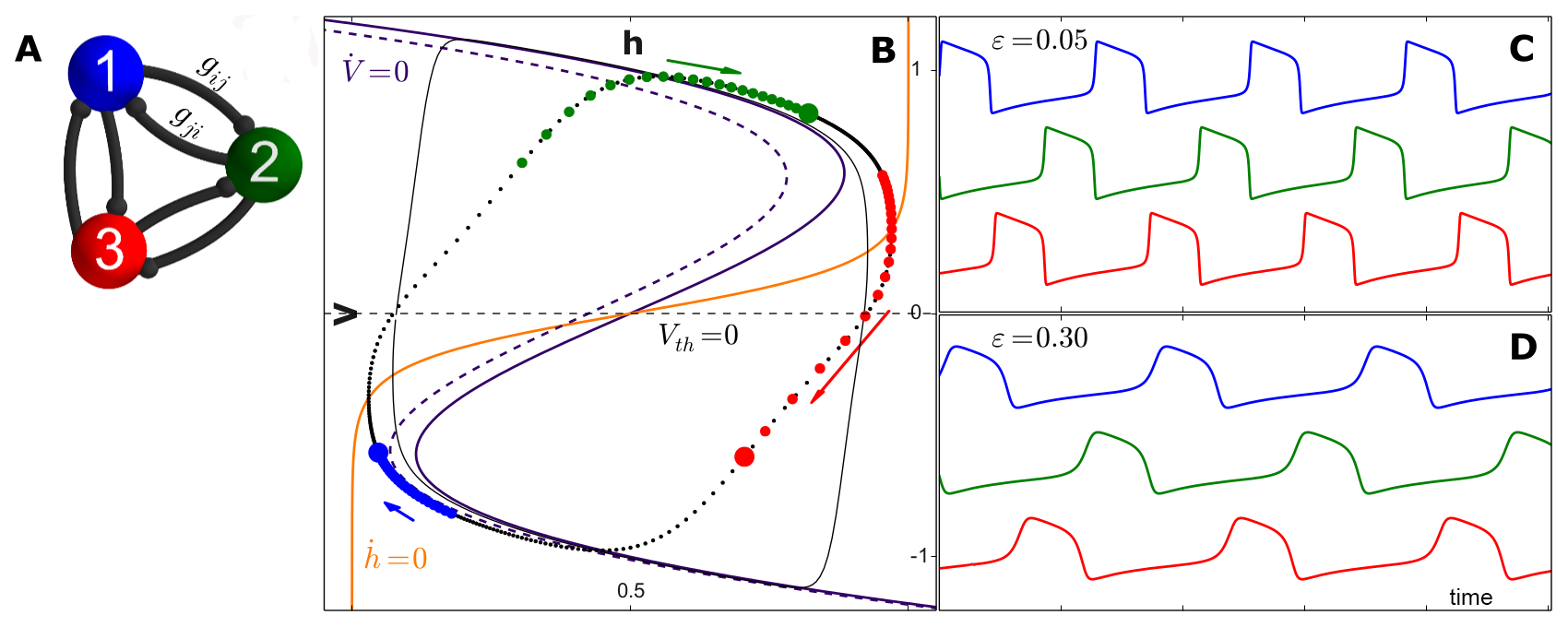}}
\caption{(A) Symmetric 3-cell circuit with inhibitory synapses. (B) The ($h,V$)-phase portrait of the three coupled cells governed by equations ~(\ref{3cellFHeq}), depicting two stable periodic orbits: relaxation-like and round-shaped (shown as gray solid and  dotted curves), at $\epsilon=0.05$ and $\epsilon=0.3$, respectively, which are  superimposed with the fast cubic nullcline (dark purple curves -- solid unperturbed  and dashed in the inhibited case) labeled as $\dot{V}=0$ and the slow sigmoidal nullcline (orange curve), $\dot{h}=0$. Blue, green and red dots on the clock-wise periodic orbit represent the time-evolution of the phases of the corresponding cells - 1, 2 and 3, coupled in the network. (C,\, D) Voltage traces generated by the network at $\epsilon=0.05$ and $\epsilon=0.3$; see the corresponding limits cycles in panel~B.
}
\label{fig:3cellmotif}
\end{center}
\end{figure*}

This paper is organized as follows: first we discuss the gFN models of neurons coupled in a network with fast inhibitory synapses.  Next, we introduce phase lags between oscillatory neurons, followed by how 2D Poincar\'e return maps for such phase lags are defined, and the correlations between the fixed points of these maps and the multiple phase-locked states of the corresponding voltage traces. We examine several representative networks to demonstrate the association between the repertoire of possible rhythmic outcomes for a given motif and its parameter space. We also discuss the role of non-local bifurcations and how these shape the borderlines separating the attraction basins in the 2D Poincar\'e map. This is followed by a discussion of some exotic motifs without any phase-locked rhythmic states. Also of our special interest is the case of the motif that opens up another possibility  -- the robust synchronous state when all three cells oscillate together. Finally, we introduce the concept of the reduced $2\theta$-neurons to build even simpler multifunctional oscillatory motifs.

\section{\label{sec:Methods} Methods}
Fitzhugh-Nagumo-like cells in biological sciences, or generic relaxation oscillators, present a mathematical generalization that captures some plain dynamical features often observed and reported in detailed HH-type models. The generalized Fitzhugh-Nagumo (gFN) model of neurons employed in this paper adds a set of extra dynamical and temporal features to  more realistically replicate biological (endogenous) bursters in isolation and, what is more important, under perturbations. We consider  a fully connected 3-cell circuit of such gFN-neurons~\cite{schwabedal2016qualitative}: 
\begin{equation}
\begin{array}{rcl}
\dot{V_i} &=& V_i-V_i^{3}-h_i+I_{app}+ \sum\limits_{j \neq i}{G_{ji}(V_j,V_i)},\\
\dot{h_i} &=& \varepsilon \left [ \dfrac{1}{1+e^{-k(V_i-V_0)}}-h_i \right ],\quad (i,j = 1,2,3).\\
\end{array}\label{3cellFHeq}
\end{equation}
Here, the state of the $i$-th neuron is described by its activity variable $V$, representing the membrane voltage, and a recovery voltage-gated variable $h$, introduced in a way similar to the Hodgkin-Huxley formalism; $\epsilon$ is the reciprocal of some time constant and regulates slow-fast dynamics ($0 < \epsilon < 1)$ in the gFN-neuron (see voltage traces in Fig.~\ref{fig:3cellmotif}C,D), while an applied current $I_{app}$ is used as a bifurcation parameter for individual cells; $V_0$ and $k$ influence the position and shape of the cubic and the sigmoidal nullclines given by $\dot V=0$ and $\dot h=0$, respectively (see Fig.~\ref{fig:3cellmotif}B). The default values for the parameters are $k=10$, $\epsilon=0.3$, and $V_{th}=0$. By construction, active ($V_i >$ 0) driving or pre-synaptic neurons slow down or repress the recovery dynamics of the driven or post-synaptic  oscillators via a fast inhibitory coupling given by $G_{ji}$, modeled using a sigmoidal coupling function employed via fast-threshold modulation (FTM) \cite{FTM}:

\begin{equation}
\begin{array}{rcl}
G_{ji}(V_j,V_i) &=& g_{ji}(V_{rev}-V_i)\Gamma(V_j),\\ 
\Gamma(V_j) &=& \dfrac{1}{1+e^{-100(V_j-V_{th})}}.
\end{array}\label{3cellGeq}
\end{equation}

The FTM-formalism can sharply differentiate the active ``on'' state of the neuron, when its voltage $V_j$ is above the synaptic threshold $V_{th} = 0$, and hence $\Gamma (V_j)=1$, from the inactive ``off'' state with $\Gamma (V_j)=0$, when $V_j<V_{th}$, provided that $\Gamma$ is stiff enough (due to the factor 100 in its equation). The strength of coupling is controlled by the maximal conductance $g_{ji}$; its default value is set to $0.001$, unless otherwise specified, to ensure weak coupling in the network. The choice of $V_{th} = 0$ and $V_{rev}=-1.5$ that makes $(V_{rev}-V_i)<0$, defines the inhibitory synapse projected from the active neuron $j$, with $\Gamma (V_j)=1$, to neuron $i$, that slows down the rate $\dot V_i$ in Eqs.~(1). Geometrically (see Fig.~\ref{fig:3cellmotif}B), the term $g_{ji}(V_{rev}-V_i)$ shifts and skews the fast nullcline $\dot V_i=0$ closer to the slow nullcline $\dot h_i=0$ of the inhibited post-synaptic neuron in the phase space. As depicted in Fig.~\ref{fig:3cellmotif}B, an individual gFN-neuron has a single unstable equilibrium state at the intersection of the fast cubic $V$-nullcline ($\dot{V}=0$) and the slow sigmoidal $h$-nullcline ($\dot{h}=0$), surrounded by a stable limit cycle  producing the relaxation oscillations shown. Note that strong inhibition or negative applied current $I_{app}$, which respectively induce temporary or permanent shifts of the $V$-nullcline, can make it cross the $h$-nullcline near the lower knee to give rise to another stable equilibrium on the lower -- hyperpolarized stable branch of the cubic nullcline, at which the inhibited neuron will rest (described further in Fig.~\ref{fig:keymechs}). If inhibition is not strong enough, it slows down the recovery or, equivalently, extends the inactive ``off'' state of the post-synaptic neuron by bringing the lower knee of the fast $V$-nullcline closer to the sigmoidal $h$-nullcline and narrowing the gap between them. This is in accordance with the famous bottle-neck effect of the saddle-node bifurcation, where the dwelling speed, estimated as $\sqrt{[\dot V]^2 + [\dot h]^2}$, decreases further the closer the nullclines approach to tangency.

\begin{figure*}[ht!]
\centering
\includegraphics[width=1.7\columnwidth]{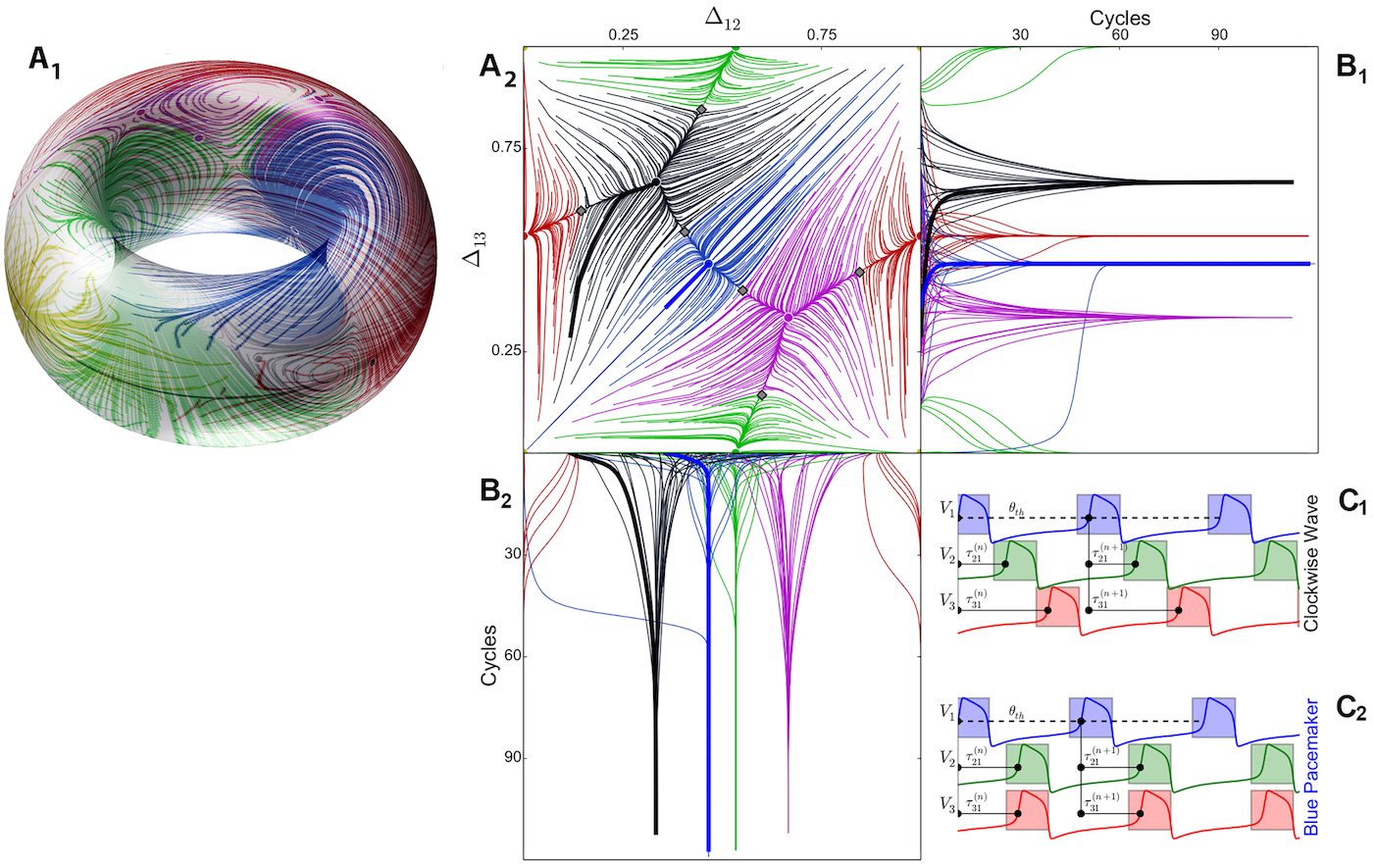}
\caption{(A$_1$) 2D torus wrapped around a Poincar\'e return map for the phase-lags between the three cells.   (A$_2$) Flattened map on a unit square, revealing 5 stable fixed points (FPs) ($\bullet$), with their color-coded attraction basins that correspond to the phase-locked states to which the phase lags $\Delta_{13}^{(n)}$  and $\Delta_{12}^{(n)}$ converge in panels B$_{1,2}$.  (C) Time-delays $\tau_{21}^{(n)}$ and  $\tau_{31}^{(n)}$ between the upstrokes in the reference blue cell~1 and in cells 2 (green) and 3 (red),  normalized over the network period, defining the phase-lags. These traces exponentially converge to multiple phase-locked rhythms such as the clockwise traveling wave or the pacemaker, corresponding to the black and blue FPs in Panel~A$_2$.}
\label{fig:phaselagreturns}
\end{figure*}

In what follows, we will show that 3-cell gFN networks can produce various stable phase-locked rhythms including the traveling waves, in which the cells fire sequentially one after the other  (see Fig.~\ref{fig:3cellmotif}C-D and  Fig.~\ref{fig:phaselagreturns}B), as well as the pacemakers, in which one cell effectively inhibits and fires in anti-phase with the remaining pair (see Fig.~\ref{fig:phaselagreturns}C). The symmetric connectivity in this network implies the coexistence of multiple rhythms that result from cyclic permutations or relabeling of the cells. In order to analyze the stability of various recurrent rhythms produced in a network, we employ the approach of Poincar\'{e} return maps.  First we introduce the notion of phase lags between the constituent cells, defined at specific events in time when the cells cross the threshold voltage from below, signaling the burst initiation. The phase lag of a cell is then defined as the delay in its burst initiation with respect to that of the reference cell 1, normalized over the bursting period. We define the $n$-th phase lags $\Delta^{(n)}_{12}$ and $\Delta^{(n)}_{13}$ of cells 2 and 3, resp., as follows:  

\begin{equation}
\Delta_{12}^{(n)} = \frac{\tau_{2}^{(n)}-\tau_{1}^{(n)}}{\tau^{(n+1)}_1-\tau^{(n)}_1}~,\quad \Delta_{13}^{(n)} = \frac{\tau_{3}^{(n)}-\tau_{1}^{(n)}}{\tau^{(n+1)}_1-\tau^{(n)}_1},  \quad \mbox{(mod~1)}, 
\label{phaselageq}
\end{equation}
where $\tau^{n}_i$ represents the time at which the $i^{th}$ cell reaches the threshold voltage, $V_{th}$=0, for the $n^{th}$ time (see Fig.~\ref{fig:phaselagreturns}B,C). The sequence of phase lags $\left \{ \Delta_{12}^{(n)}, \, \Delta_{13}^{(n)} \right \} $, defined for values between $0$ and $1$, gives the forward phase trajectory on a 2D torus (Fig.~\ref{fig:phaselagreturns}A). The specific phase lag values $0$ (or $1$) and $0.5$ represent in-phase and anti-phase relationships, respectively, with the reference cell~1. We examine the 2D phase space of the Poincar\'{e} return maps (Fig.~\ref{fig:phaselagreturns}A) of the 3-cell networks, given by the phase lags $\Delta_{12}$ and $\Delta_{13}$, by initiating multiple trajectories with different initial phase lags (on a grid of size $50 \times 50$), and by following their evolutions for a large number of cycles. As we compute long traces of firing activity of the circuit and evaluate the corresponding phase lag trajectories, these phase lags eventually converge to some attractor which can be  a fixed point of the 2D map (with steady coordinates $\Delta^*_{12}$ and $\Delta^*_{13}$ in (\ref{phaselageq})), which implies the existence of a stable rhythmic pattern in the circuit with phase-locked bursting between the cells. All the phase trajectories that converge to the same fixed point are marked by the same color and reveal the basin of attraction for the corresponding stable rhythm. The rhythmic characteristics of a CPG can be identified by analyzing the phase space of the corresponding Poincar\'{e} map, such as Fig.~\ref{fig:phaselagreturns}A which reveals the existence of penta-stability in the circuit, composed of three pacemakers (blue - Fig.~\ref{fig:phaselagreturns}C$_2$, green and red) and two traveling waves (clockwise -  Fig.~\ref{fig:phaselagreturns}C$_1$, and anti-clockwise). We depict stable fixed points with colored dots, saddles with gray diamonds, and unstable fixed points with white dots in the 2D maps. The phase lag values ($\Delta_{12}$, $\Delta_{13}$) at the fixed points (FPs) corresponding to the blue, green and red pacemakers (PMs) are given by $(0.5, 0.5)$, $(0.5, 0)$ and $(0, 0.5)$, respectively, while those for the clockwise (black) and anti-clockwise (purple) traveling-waves (TWs) are given by $(0.33, 0.67)$ and $(0.67, 0.33)$, respectively.

Alternatively, it is also possible that the forward phase-lag trajectories $\left \{ \Delta_{12}^{(n)}, \, \Delta_{13}^{(n)} \right \} $, could converge to attractors other than FPs, as demonstrated in Fig.~\ref{fig:doublerel}D\textprime{}, Fig.~\ref{fig:doubleesc}F, and Figs.~\ref{fig:kingesc}A,B and E, or the map has no attractors at all as in Fig.~\ref{fig:ErgodicTorus}. 
One can also trace down the backward trajectories, as shown in Fig.~\ref{fig:ErgodicTorus}, to effectively depict repelling fixed points and unstable phase-slipping orbits. 

Numerical integration of the trajectories is performed using the fourth order Runge-Kutta method, with a fixed step size. Computation of voltage and phase lag trajectories across multiple initial conditions is parallelized on a Tesla K40 GPU using CUDA, while visualizations are performed in Python. The open source software toolkit developed is available at \url{https://github.com/jusjusjus/Motiftoolbox} \cite{motiftoolbox}. GPU parallelization allows the construction of typical phase sweeps such as one shown in Fig.~\ref{fig:phaselagreturns}A within just a few seconds.

\begin{figure*}[ht!]
\centering
\includegraphics[width=1.6\columnwidth]{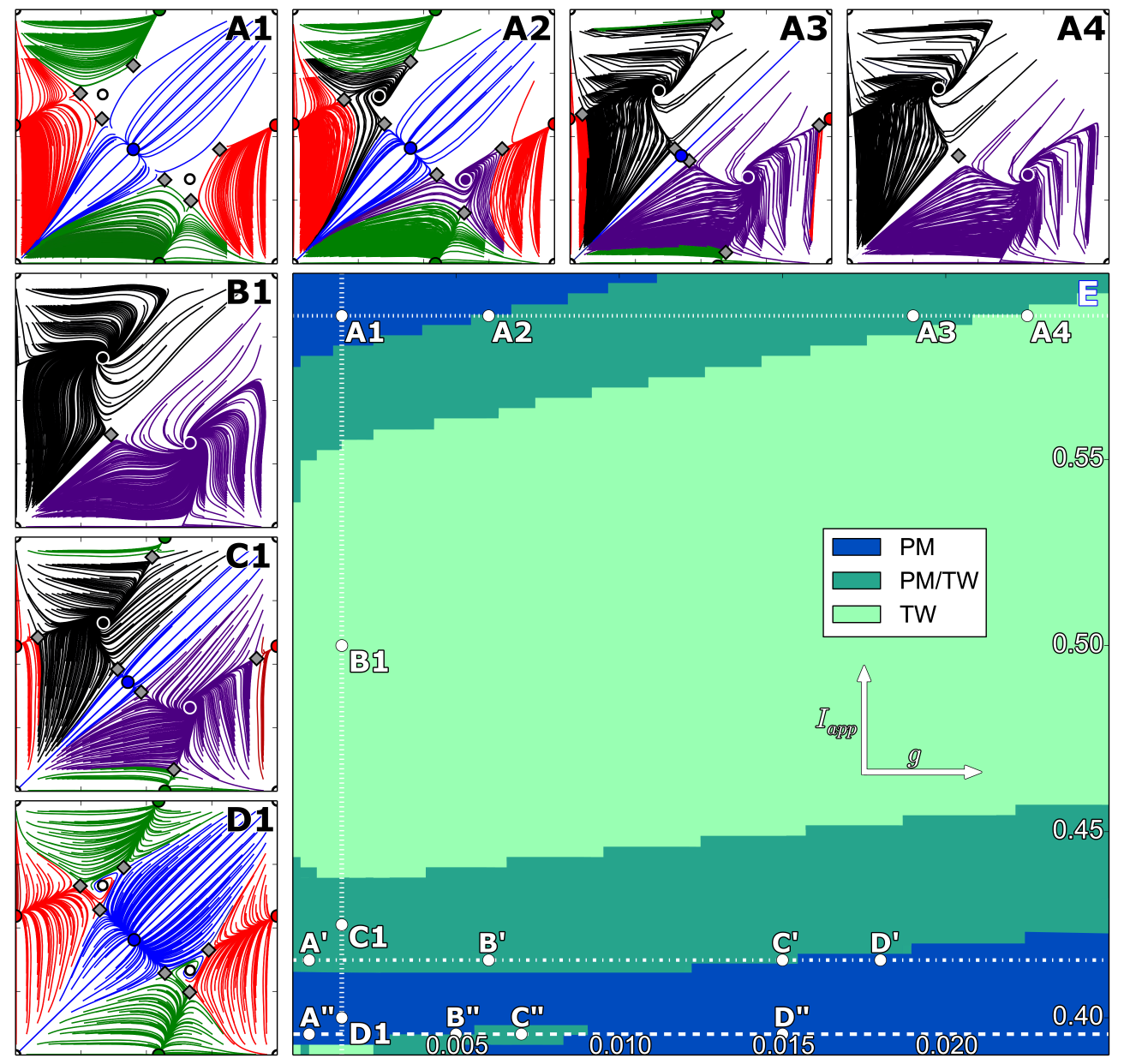}
\caption{The ($g,\, I_{app}$)-bifurcation diagram in (E) for the fully symmetric 3-cell motif (in Fig.~\ref{fig:3cellmotif}A) has several (color-coded) regions where the network can produce only pace-maker (PM) rhythms, or only traveling wave (TW) rhythms, or both PMs/TWs. Panels~A1--A4 give the snapshots of the Poincar\'{e} return maps for the network due to the escape mechanism as they are sampled at the parameter values (white dots) along the top horizontal line in panel (E). As the coupling strength $g$ increases, the map with  3 stable FPs: blue PM at (0.5, 0.5), green PM at (0.5, 0) and red PM at (0, 0.5) in (A1), gains two more stable FPs: black TW at (0.33, 0.67) and purple TW (0.67, 0.33) in (A2). The growing TW and shrinking PM attraction basins are separated by the separatrices of the saddles (grey $\diamond$)  in (A3). Each stable PM become a saddle through a pitch-fork bifurcation at larger $g$-values, after merging with two nearby saddles in (A4). Panels (A1-D1) snapshot several maps as the network transitions from the escape to the release mechanism as $I_{app}$  is decreased (along the left vertical dotted lines in E): from 3 stable PMs in (A1), to two stable TWs in (B1), to mixed dynamics with both PM/TWs in (C1), and back to 3 stable PMs in (D1). Parameters for (A1-A4): $I_{app} = 0.5886$, $g = (0.0015, 0.006, 0.019, 0.0225)$; parameters for (B1-D1): $g = 0.0015$, $I_{app} = (0.493, 0.419, 0.393)$. The return maps sampled along the horizontal dashed lines A'-D' and A"-D"  at the bottom of panel (E) are presented in Figs.~\ref{fig:relsymI42} and \ref{fig:relsymI39} for the networks obeying the release mechanism.
}
\label{fig:relsymgrid}
\end{figure*}

\begin{figure}[hbt!]
\begin{center}
\resizebox*{0.6\columnwidth}{!}
{\includegraphics{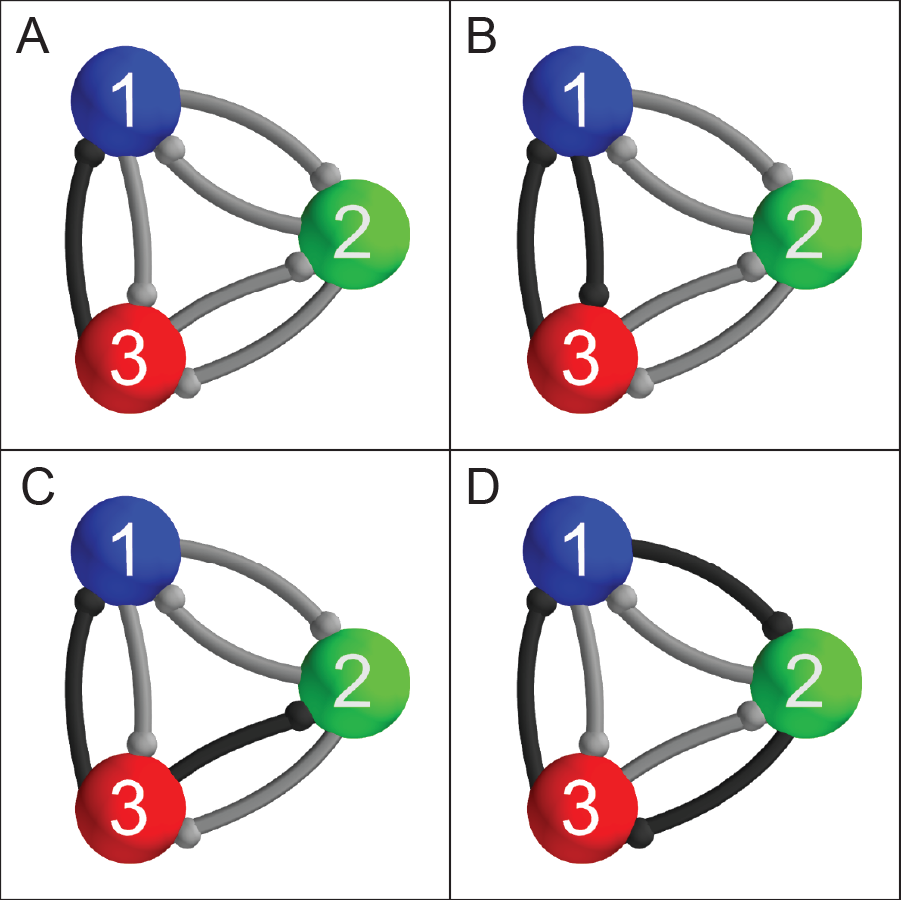}}
\caption{Key asymmetric configurations analyzed in this study: (A) Mono-biased motif where only the synapse $g_{31}$ is manipulated, (B) Double-biased motif, in which the reciprocal connections between cells 1 and 3 ($g_{31}$ and $g_{13}$) are manipulated equally, (C) Driver-biased motif, in which both the outgoing connections from cell 3 ($g_{31}$ and $g_{32}$) are affected equally, and (D) Clockwise-biased motif, in which all the clockwise connections ($g_{12}$, $g_{23}$, and $g_{31}$) are affected equally.
}
\label{fig:AsymMotifs}
\end{center}
\end{figure}

With detailed simulations of the Poincar\'{e} return maps for phase lags, we can visualize and analyze their stable and unstable fixed points (FPs), and other limiting sets such as invariant circles (ICs). We can also detect various bifurcations including homoclinic and heteroclinic, and therefore can identify multiple possible rhythmic patterns generated by the neural circuits. With changes in the bifurcation parameters of the network, the constructed phase-leg sequences also vary, thus allowing us to determine the basins of the coexisting attractors and to reveal bifurcations through which FPs can emerge, disappear or lose their stability. By varying two bifurcation parameters of the circuit (typically, the external drive $I_{app}$ and the coupling strength $g$) and identifying all the permissible rhythmic states, we construct bi-parametric sweeps such as one shown in Fig.~\ref{fig:relsymgrid}E. The given bifurcation diagram frames the neighborhoods of the release and the escape mechanisms (described later) in the parameter plane for the symmetric network (depicted in Fig.~\ref{fig:3cellmotif}A) over a wide behavioral range of the individual neurons. We can characterize the parametric regions whose phase-lag return maps contain only the pacemakers (PM), only the traveling waves (TW), or a combination of both PM/TW. The return maps shown in panels (A1-A4) and (A1-D1) in Fig.~\ref{fig:relsymgrid} reveal the sequence of bifurcations that stable rhythms undergo near the borderlines. Such bifurcation diagrams have proven useful for studying the dynamics of small CPG-circuits and other nonlinear systems  \cite{pusuluri2019computational, pusuluri2019symbolic, Pusuluri2020, pusuluri2017unraveling, pusuluri2018homoclinic}. In addition to the fully symmetric motif in Fig.~\ref{fig:3cellmotif}A, we perform a detailed bifurcation analysis of some other key asymmetric motifs (see Fig.~\ref{fig:AsymMotifs}). They include (1) Mono-biased motif, in which only a single synaptic connection is varied while all others are held constant; (2) Double-biased motif, in which the reciprocal connections between two cells are manipulated equally, while all others are held constant; (3) Biased-driver motif, in which both the outgoing connections from one cell are varied identically, while the rest are held constant; and (4) Clockwise-biased motif, in which all the clockwise connections are changed simultaneously, while the anti-clockwise connections remain fixed. This is followed by a brief description of another asymmetric network configuration (see Fig.~\ref{fig:ErgodicTorus}) without any phase-locked rhythmic states, but only quasi-periodic phase slipping or an ergodic flow with neither FPs nor ICs. 

\section{Results and Discussion}
\begin{figure}[hbt!]
\begin{center}
\resizebox*{0.99\columnwidth}{!}{\includegraphics{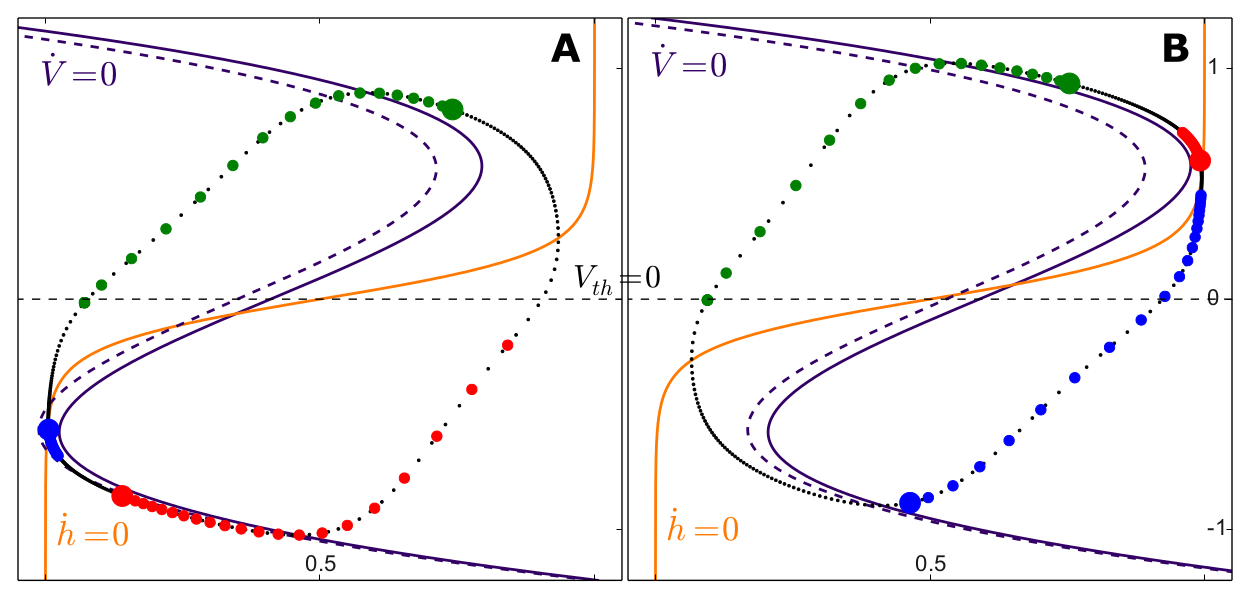}}
\caption{ Phase portraits of the gFN-neuron given by Eqs.~(\ref{3cellFHeq}), demonstrating the release $(A)$ and escape $(B)$ mechanisms of rhythmogenesis in the 3-cell circuit.  The clusters of the colored dots representing the phases of the coupled cells on the periodic orbit, at the lower-left and upper-right knees near the tangencies of the fast cubic and slow sigmoidal nullclines $\dot V=0$ and $\dot h=0$, respectively, are indicative of the stagnation due to the saddle-node bottleneck effect underlying the release and escape mechanisms. The solid (unperturbed) and dashed (perturbed) purple curves show the shifts in the $V$-nullclines for the pre- and post-synaptic cells, respectively. The intersection of the fast dashed $V$-nullcline in (A) with  the slow $h$-nullcline near the left knee,  corresponds to a newly formed stable fixed point that makes the post-synaptic cell ``hard-locked'' by synaptic inhibition. In case (B), the post-synaptic cell escapes from the trap (hard or soft) of depolarization near the upper knee, after an inhibitory perturbation shifts $\dot V=0$ leftward to create a gap between the nullclines, allowing the cell(s) to fall down on to the hyperpolarized branch, to start a new cycle of revolution.   
}
\label{fig:keymechs}
\end{center}
\end{figure}

The terms ``release'' and ``escape'' mechanisms referring to anti-phase oscillations in 2-cell networks  were first introduced by Wang and Rinzel \citep{wang1992alternating}. By construction, the release mechanism requires that isolated cells are intrinsically bursting, while the escape mechanism requires the cell to wait at the depolarized quiescent state, which can be also associated with the tonic-spiking state in the case of the ``spike-less'' gFN-burster. The initial state of the gFN-neuron depends on the value of the applied current stimulus $I_{app}$ in Eqs~.(1), which horizontally  shifts the position of the fast $V$-nullcline given by $\dot V=0$, relative to that of the slow $h$-nullcline given by $\dot h=0$, see Figs.~2,\ref{fig:keymechs}. In this study we examine how these mechanisms influence the dynamical behavior, rhythmic outputs, and bifurcations in the 3-cell motifs with several distinct circuitries presented in Fig.~\ref{fig:3cellmotif}$A$ and in  Fig.~\ref{fig:AsymMotifs}, using the methodology outlined in Section~\ref{sec:Methods}.

A weakly coupled network obeys the release mechanism when the slow $h$-nullcline lies close enough to the lower knee of the fast cubic $V$-nullcline in the phase space of an intrinsic HH burster in Fig.~\ref{fig:HH} or in the ($h,\,V$)-plane of the gFN-neuron in Fig.~\ref{fig:keymechs}A. As the value of the $I_{app}$-parameter decreases, or alternatively, if a cell receives an inhibitory synaptic current from its pre-synaptic cell(s), its $V$-nullcline shifts horizontally leftwards in the ($h,\,V$)-plane. In the release case, this can lead to the occurence of tangency of both the nullclines near the lower knee of the $V$-nullcline. Further strengthening of inhibition causes the nullclines to cross locally twice, giving rise to a new stable equilibrium state on the lower hyperpolarized branch on the $V$-nullcline, that emerges through a saddle-node bifurcation. This ceases oscillations in the cell, as it remains effectively hard-locked in the hyperpolarized state. Otherwise, if the inhibition is not strong enough to cause the saddle-node bifurcation, it merely narrows the gap between the nullclines which makes the cell slow down due to the bottleneck effect preceding the saddle-node bifurcation. After the presynaptic cell(s) traverses its active on-state and becomes inactive on its lower hyperpolarized branch of the fast $V$-nullcline below the synaptic threshold, the post-synaptic cell is released from inhibition and its $V$-nullcline shifts rightward back to its original position. This eliminates the stable hyperpolarized equilibrium state, so the given cell can start a new oscillatory cycle and during its active phase, can in turn play the opposite role of inhibiting its post-synaptic cells in the network.

  The escape mechanism requires that (i) one of the unperturbed gFN-cells, say the red cell 3, is initially locked at the stable depolarized equilibrium state at the transverse intersection of the slow  $h$-nullcline  and the fast $V$-nullcline near its upper right knee, as shown Fig.~\ref{fig:keymechs}B. The other condition (ii) is that meanwhile the other cell(s) must transition along the perturbed hyperpolarized branch of the $V$-nullcline till either post-synaptic cell, say the green cell 2, reaches the lower knee from where it jumps up, switching from the ``off'' to ``on'' states. As soon as its voltage goes over the synaptic threshold, cell~2 results in a fast inhibitory current that shifts the $V$-nullcline of cell 3 leftwards, away from the $h$-nullcline, and eliminates the stable equilibrium state, so that cell 3 jumps down on to the inhibited hyperpolarized branch of the $V$-nullcline, and so forth. 
  
   Below, we will show how inhibitory 3-cell networks can employ the two generic mechanisms to synergistically orchestrate the shifts in the positions of the nullclines of post-synaptic neurons, so that the constituent cells can alternate cyclically between their active and inactive states, resulting in several stable rhythmic outcomes. We reiterate that we only consider weakly coupled networks here, to ensure the visual continuity of the Poincar\'e return maps for the phase lags. Our choice of the sigmoidal shape for the slow $h$-nullcline ensures that the system can exploit the bottleneck effect of saddle-node bifurcations to produce a variety of rhythmic outcomes. Increasing coupling strength causes faster non-smooth convergence and hard-locking to stable FPs in the maps, corresponding to phase-locked rhythms; see Fig.~\ref{fig:relsymI39} illustrating the effect of increasing inhibition in the symmetric motif.        

  In the following sections, using Poincar\'{e} return maps and bifurcation diagrams, we will demonstrate the onset of various rhythms and their transitions in the fully symmetrical system, as we vary control parameters that can also be manipulated in neurophysiological experiments. We will then show the behavioral ranges and transitions occurring in various asymmetrical network configurations, otherwise impossible in the fully symmetric system. This is followed by a brief discussion of the stable synchronous state with all three cells oscillating together, as well as a special network configuration without any phase-locked rhythms.

\subsection{Symmetric motif}

Figure~\ref{fig:relsymgrid}E represents the so-called bi-parametric sweep, or the bifurcation diagram, of the fully symmetric 3-cell motif depicted in Fig.~\ref{fig:3cellmotif}A as two parameters: the coupling strength $g$ of all six inhibitory synapses and the applied current $I_{app}$, are varied. One can see that the parameter space has three color-coded regions where the network produces either three stable pacemakers (PM), or two traveling waves (TW), or five coexisting rhythms: 3 PMs and 2 TWs. 

\begin{figure}[h]
\begin{center}
\resizebox*{0.99\columnwidth}{!}
{\includegraphics{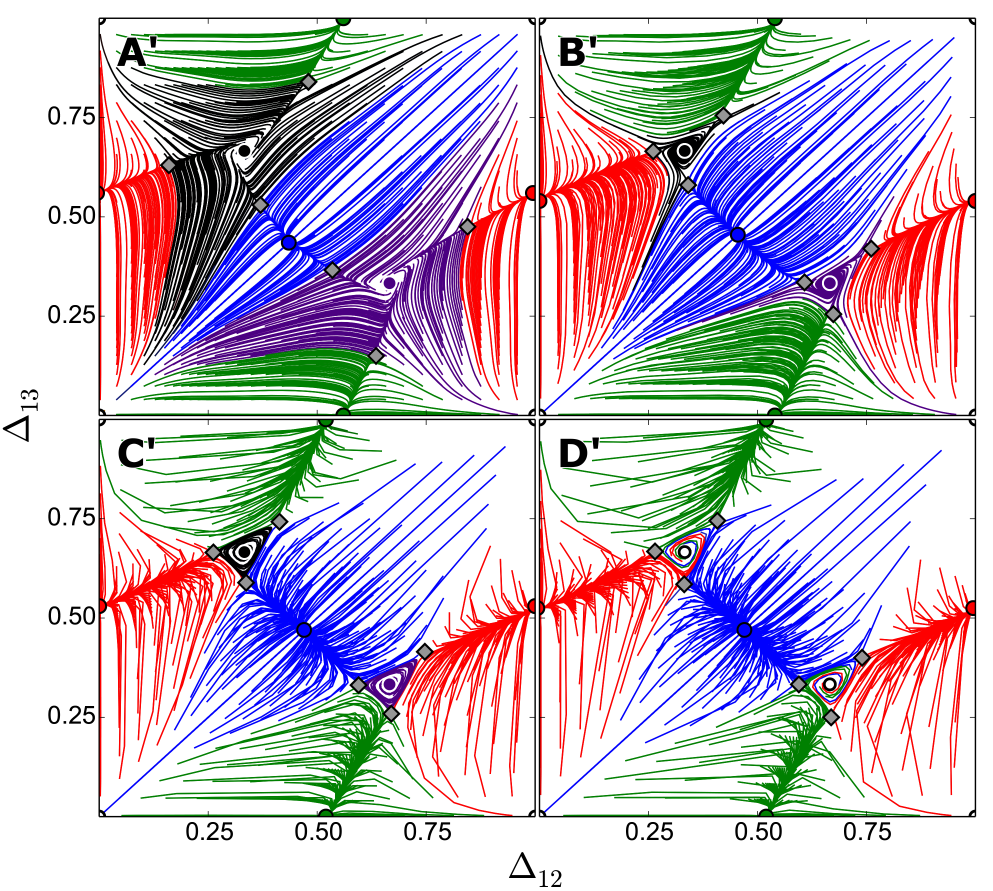}}
\caption{Transformations in the 2D return maps corresponding to the fully symmetric networks with the release mechanism as the inhibitory coupling $g$ is increased along the dashed white line (A\textprime{}-D\textprime{}) in the bifurcation diagram in Fig.~\ref{fig:relsymgrid}(E). The TW attraction basins decrease in size, while the PM basins increase from A\textprime{} through C\textprime{}. The TWs lose stability in (D\textprime{}) through  a secondary Andronov-Hopf/torus bifurcation. Less smooth trajectories are indicative of fast convergence to the attractors in the network, at greater coupling strengths.  Parameters: $I_{app} = 0.4155$, $g = (0.0005, 0.006, 0.015, 0.018)$}
\label{fig:relsymI42}
\end{center}
\end{figure}

The corresponding 2D Poincar\'e maps for the phase lags, depicted in Figs.~\ref{fig:relsymgrid}A1--A4, demonstrate the transitions and bifurcations in the network due the escape mechanism as the inhibitory coupling is increased. The initial gap between the slow $h$-nullcline and the upper knee of the fast $V$-nullcline is small enough so that the modulating bottleneck effect makes either cell linger longer in the active on-phase near the upper knee, while the other two cells transition through the inactive off-phase, thereby promoting pacemaker rhythms corresponding to the three stable FPs: blue at $(0.5, 0.5)$, green at $(0.5, 0)$ and red $(0, 0.5)$. As the synaptic coupling $g$ is increased in strength, the gap between the nullclines of the post-synaptic cells widens (see Fig.~\ref{fig:keymechs}B), thus weakening the bottleneck effect so that the circular motion on the limit cycles in the phase plane becomes more uniform. As we increase $g$ further, the unstable TWs at $(0.33, 0.67)$ and $(0.67, 0.33)$ become stable through a secondary Andronov-Hopf or torus bifurcation (Fig.~\ref{fig:relsymgrid}A2), with the TW attraction basins gradually increasing, while those of the PMs diminishing in size (Fig.~\ref{fig:relsymgrid}A3). Note that there are an even number of saddle FPs (labelled by grey $\diamond$s ) in the maps: their (separatrices') role is to separate the attraction basins of the coexisting stable FPs. With further increase in $g$, a pair of nearby saddles approach each stable PM fixed point and merge with it through a pitch-fork bifurcation. The three new saddles now equally partition the attraction basins of the two remaining TWs in Fig.~\ref{fig:relsymgrid}A4.  
\begin{figure}[h]
\begin{center}
\resizebox*{0.99\columnwidth}{!}
{\includegraphics{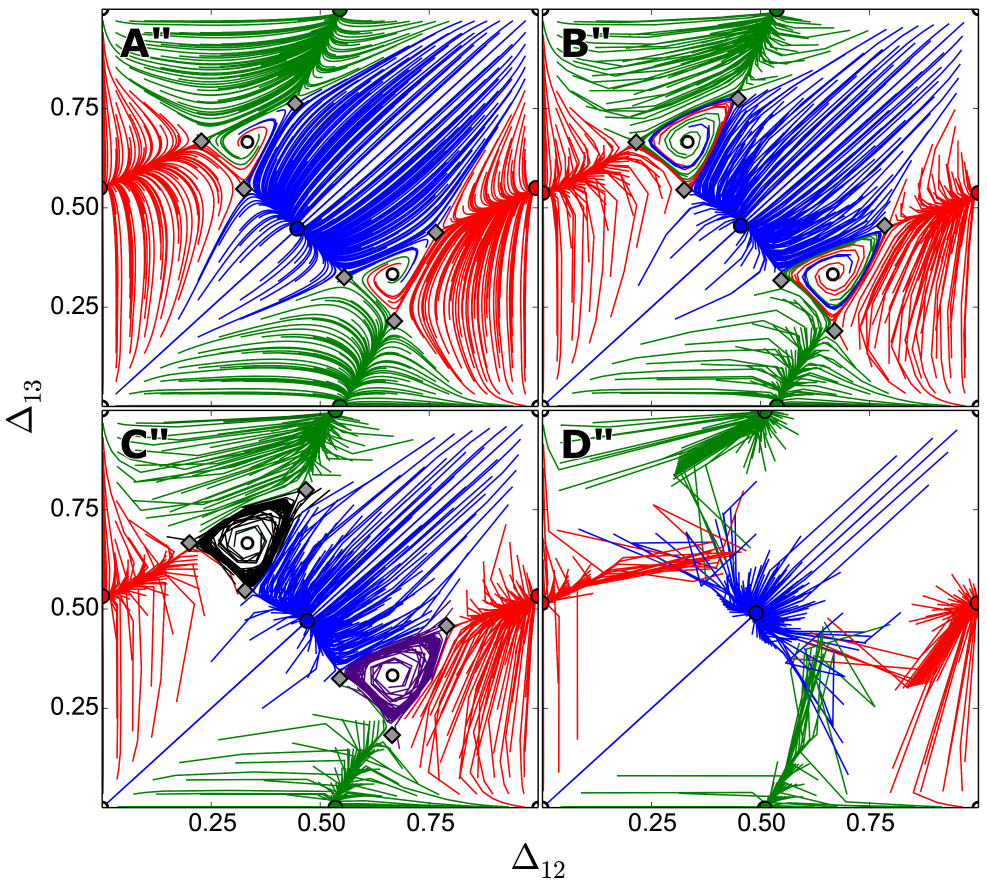}}
\caption{Poincar\'{e} return maps corresponding to the fully symmetric 3-cell motif with another example of the release mechanism for the parameter values sampled along the line (A\textprime\textprime{}-D\textprime\textprime{}) in the bifurcation diagram Fig.~\ref{fig:relsymgrid}(E). As $g$ is increased, the unstable TWs in (A\textprime\textprime{}-B\textprime\textprime{}), become stable in Panel~C\textprime\textprime{} through a torus bifurcation. The TWs again lose stability with a further increase in $g$, leading to hard-locking in the system that results in the quick and jagged convergence to the three PMs in (D\textprime\textprime{}). Parameters: $I_{app} = 0.3956$, $g = (0.0005, 0.005, 0.007, 0.015)$}
\label{fig:relsymI39}
\end{center}
\end{figure}

The maps depicted in Figs.~\ref{fig:relsymgrid}A1--D1 illustrate the bifurcation stages as the symmetric 3-cell motif transitions from the escape to the release mechanism (Fig.~\ref{fig:keymechs}). This occurs as $I_{app}$ is decreased along the vertical dotted line in the bifurcation diagram in Fig.~\ref{fig:relsymgrid}(E). Decreasing $I_{app}$ shifts the fast $V$-nullcline leftwards and therefore gradually increases the gap between its upper knee and the slow $h$-nullcline, while simultaneously decreasing the gap near the lower knee of the $V$-nullcline. As seen in the corresponding return maps in Fig.~\ref{fig:relsymgrid}, the network transitions from stage A1: a motif producing only three stable PMs; to stage B1: a motif producing only two stable TWs, after the pitch-fork and torus bifurcations; to stage C1: a motif with co-existing TWs and PMs (as PMs re-emerge after reverse pitch-fork bifurcations). Finally, reverse torus bifurcations make the TWs unstable again and restore the motif with only three stable PMs at stage D1. This is due to the dominating bottleneck effect near the lower knee, that substantially amplifies the slow-fast separation in the rate of circulation along the stable limit cycles in the ($h,\,V$)-plane. 

Figure~\ref{fig:relsymI42} shows the transitions occurring based on the release mechanism, as the synaptic inhibition is increased in the 3-cell network. The initial gap between the slow $h$-nullcline and the lower knee of the fast $V$-nullcline is chosen so that both stable PM and TW rhythms can coexist for the given coupling strength. As the synaptic inhibition $g$ is gradually increased, this gap narrows, thereby slowing the progression of the post-synaptic cells near the lower fold (dwelling time here is inversely proportional to the square root of the size of the gap between the nullclines). This makes the pacemaker activity dominant in the network. Further increase in $g$ can lead to hard-locking, as the nullclines intersect at the lower branch, after crossing the knee. This results in the gradual shrinking of the attraction basins of TWs, and a corresponding increase in those of PMs, as seen in Fig.~\ref{fig:relsymI42}A\textprime{}--C\textprime{}. TWs finally become unstable through a secondary Andronov-Hopf/torus bifurcation in Fig.~\ref{fig:relsymI42}D\textprime{}, and hence not observable in the network anymore. 

Let us now consider another bifurcation pathway in the bifurcation diagram in Fig.~\ref{fig:relsymI42}(E), for a lower value of $I_{app}=0.3956$, and therefore displaying a more pronounced release mechanism in the network. The return maps with increasing synaptic inhibition are presented in Fig.~\ref{fig:relsymI39}. As $g$ is gradually increased, the network bifurcates from the dominant PM case in Figs.~\ref{fig:relsymI39}A\textprime\textprime{}--B\textprime\textprime{}, to a configuration that supports both PMs and TWs in Fig.~\ref{fig:relsymI39}C\textprime{\textprime{}, following torus bifurcations. A further increase in the value of $g$ leads to hard-locking in Fig.~\ref{fig:relsymI39}D\textprime\textprime{}, due to the tangency or the intersection of the nullclines near the lower knee, for the temporarily inhibited post-synaptic cells. This can be seen in the jagged phase-lag trajectories in the return map, that quickly converge to the pacemaker fixed points. 

Finally, we can conclude that increasing coupling strength leads to pronounced pacemaker behaviors in the networks featuring the release mechanism. In contrary, traveling waves become more dominant at stronger coupling values in networks featuring the escape mechanism. This means that symmetry {\it per se} is insufficient to predict {\rm a priori} what rhythmic outcomes a given network can produce, without knowing the qualitative mechanisms of rhythmogenesis, escape or release, and the quantitative strength of synaptic connectivity. This knowledge is vital to make testable predictions regarding possible dynamics in various biological systems of coupled oscillators and neurophysiological experiments with CPG circuits.

\subsection{Mono-biased motif}

\begin{figure}[h!]
\begin{center}
\resizebox*{0.9\columnwidth}{!}
{\includegraphics{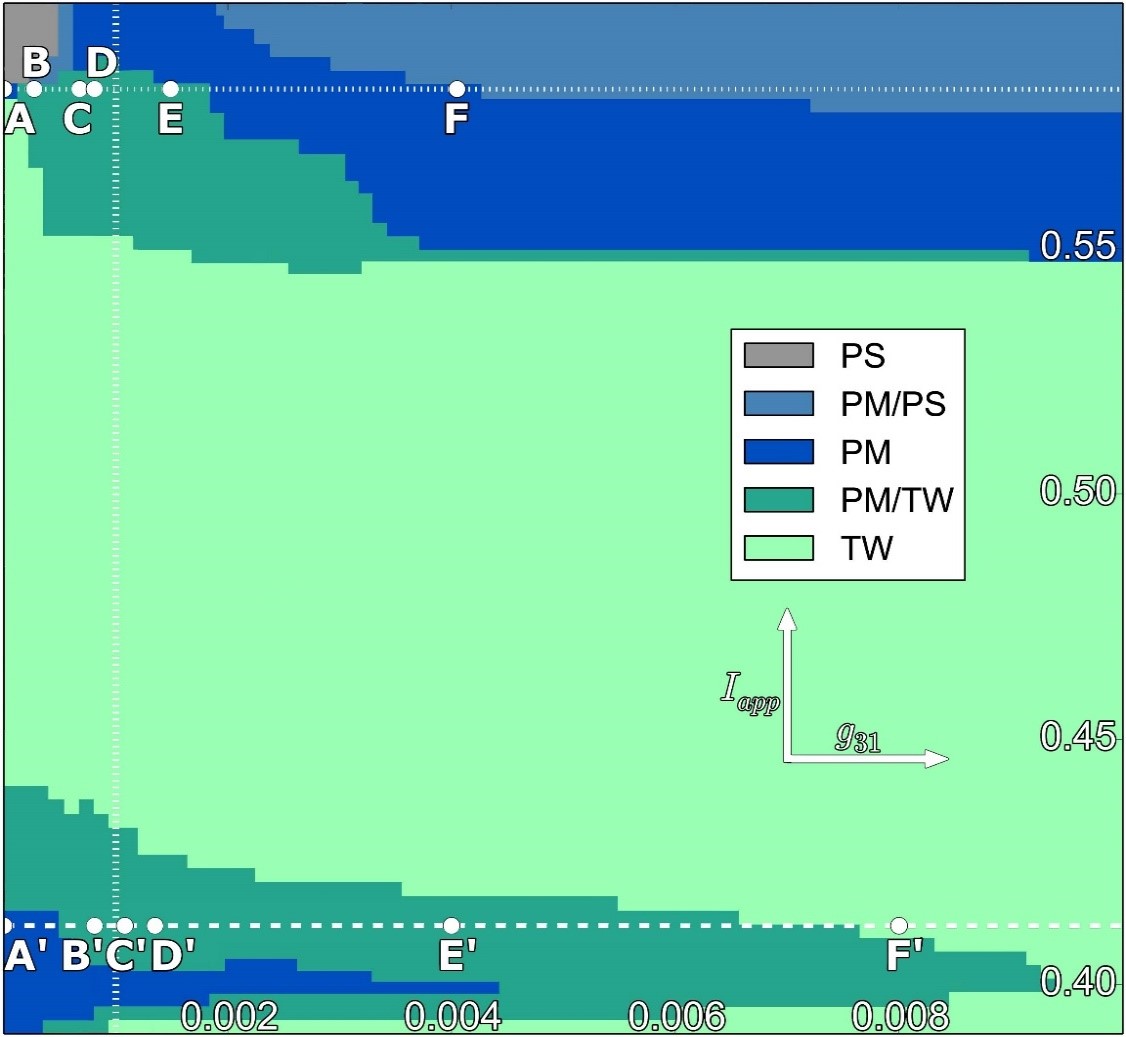}}
\caption{$(g_{31}, \,I_{app})$-bifurcation diagram of the mono-biased motif presented in Fig.~\ref{fig:AsymMotifs}A shows five distinct regions corresponding to  PM, TW, and PS rhythms along with the combinations PM/PS and PM/TW. Transitions between these regions are governed by saddle-node (SN) bifurcations that eliminate or restore FPs in the return map. The points A-F and A\textprime{}-F\textprime{} highlighted near the top dotted and bottom dashed lines, respectively, indicate the parameter values used for the return maps with the escape and release mechanisms elaborated in Fig.~\ref{fig:singrel} and Fig.~\ref{fig:singesc}, respectively. The vertical line given by $g_{31} = g_{\rm all} = 0.001$ corresponds to the fully symmetric network where all $g$-values are identical.}
\label{fig:bifsingle}
\end{center}
\end{figure}

\begin{figure}[h!]
\begin{center}
\resizebox*{0.99\columnwidth}{!}
{\includegraphics{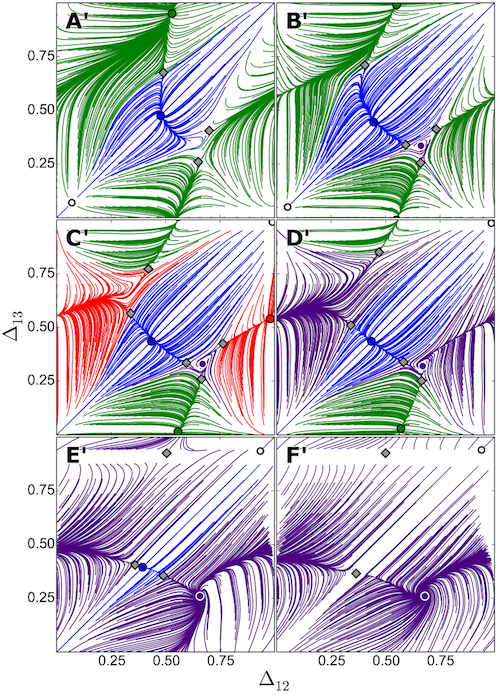}}
\caption{Poincar\'{e} return maps corresponding to the mono-biased motif with the release mechanism and their evolution as $g_{31}$ gradually increases while the remaining synaptic strengths ($g_{\rm all}$) are held constant. When $g_{31}=0$ in Panel A\textprime{}, the green PM (stable FP at (0.5,0)) dominates the dynamics, coexisting with the blue PM (FP at (0.5,0.5)) having a smaller attraction basin.  Increasing $g_{31}$ leads to the formation of the purple TW pattern (stable FP at (0.6,0.3)) through a saddle-node bifurcation, while the blue basin increases in Panel B\textprime{}, followed by the appearance of the red PM in (C\textprime{}) due to a saddle-node bifurcation, after the motif partially restores the anti-clockwise symmetry. Increasing $g_{31}$ further leads to a single dominant anti-clockwise purple TW rhythm in the network, after all other attracting FPs vanish via a series of saddle-node bifurcations in  Panels D\textprime{}-F\textprime{}. Parameters: $I_{app}$=0.412, $g_{\rm all}$=0.001, $g_{31}$=(0, 0.00081, 0.00108, 0.00135, 0.004, 0.008).}
\label{fig:singrel}
\end{center}
\end{figure}

\begin{figure}[h!]
\begin{center}
\resizebox*{0.999\columnwidth}{!}
{\includegraphics{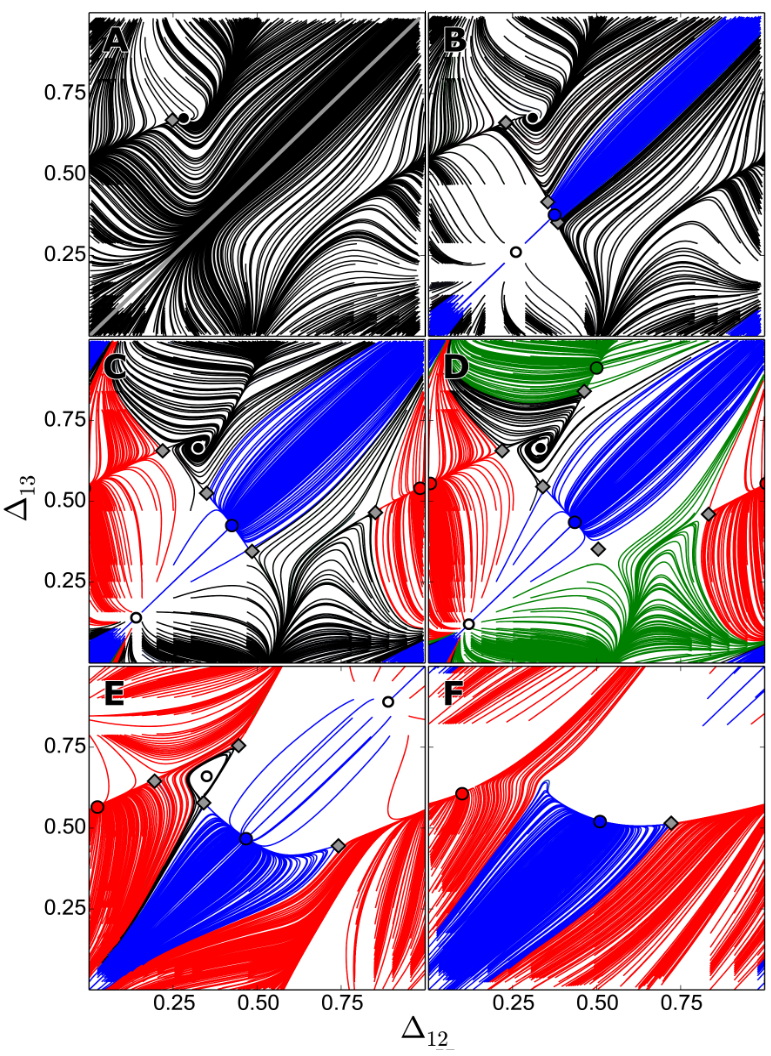}}
\caption{Poincar\'{e} return maps corresponding to the mono-biased motif with the escape mechanism as $g_{31}$ gradually increases. At $g_{31}=0$, there is a single dominant clockwise (black) TW FP in Panel A;  this is elaborated further in Fig.~\ref{fig:Black4Panels} and Fig.~\ref{fig:singEscPhaseSlip4Panels}. (B) As $g_{31}$ increases, the unstable IC (the blue line in Fig.~\ref{fig:Black4Panels} ) first undergoes a reverse homoclinic saddle-node bifurcation, giving rise to a repelling fixed point (white dot) and a saddle, which then undergoes a pitch-fork bifurcation that makes it stable -- the blue PM, with two additional saddles. The red and green PMs then emerge following additional saddle-node bifurcations, see panels C and D. The green PM disappears through a saddle-node bifurcation, while the black TW at (0.3,\,0.6) becomes repelling via a torus bifurcation, giving rise to a stable invariant circle (IC) in Panel~E, and finally gets annihilated after merging with a nearby saddle. Parameters: $I_{app}=0.5825$, $g_{\rm all}$=0.001, $g_{31}$=(0, 0.00027, 0.000676, 0.00081, 0.00149, 0.00405)}
\label{fig:singesc}
\end{center}
\end{figure}

\begin{figure}[h!]
\begin{center}
\resizebox*{0.99\columnwidth}{!}
{\includegraphics{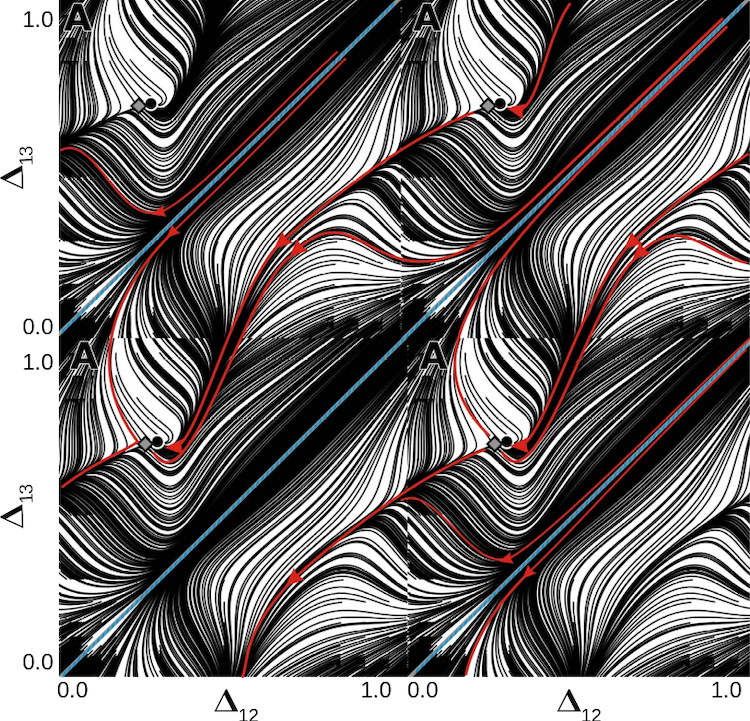}}
\caption{Four identical panels from Fig.~\ref{fig:singesc}A stitched together for a better understanding of the trajectories wrapping around the phase torus. The network is dominated by a single black TW around (0.3,\,0.6). The saddle, located in a close proximity, causes two trajectories from close initial conditions to traverse different paths (red lines) to converge to the same FP. Shown in blue is the repelling invariant circle (IC), see also Fig.~\ref{fig:singEscPhaseSlip4Panels}. Parameters: $I_{app} = 0.5825$, $g_{\rm all} = 0.001$ except $g_{31} = 0$. 
}
\label{fig:Black4Panels}
\end{center}
\end{figure}

\begin{figure}[h!]
\begin{center}
\resizebox*{0.99\columnwidth}{!}
{\includegraphics{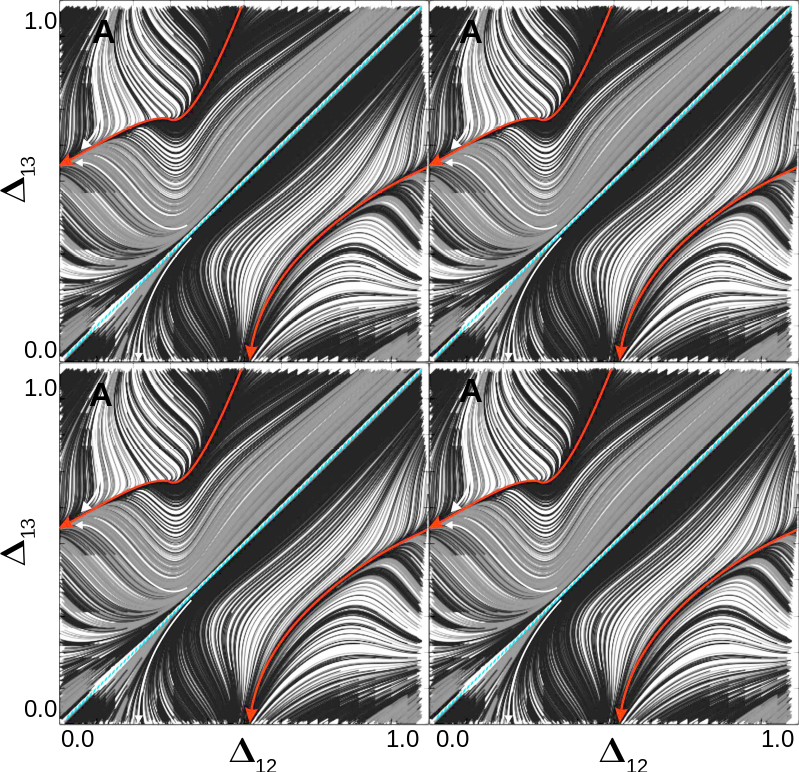}}
\caption{Four identical panels stitched together to better visualize the trajectories wrapping around the phase torus, as a stable PS pattern emerges following the disappearance of the black TW and the nearby saddle of Fig.~\ref{fig:Black4Panels}, through a homoclinic saddle-node bifurcation. The network has no phase-locked rhythms and all trajectories converge on to the stable invariant curve (red). The blue line marks an unstable invariant curve. Parameters: $I_{app} = 0.5875$, $g_{\rm all} = 0.001$ except $g_{31} = 0$.}
\label{fig:singEscPhaseSlip4Panels}
\end{center}
\end{figure}

We will now investigate the effect of a single asymmetric connection within an otherwise fully-symmetric, weakly-coupled system (see Fig.~\ref{fig:AsymMotifs}A). While we focus on asymmetric increase or decrease in the strength of a single connection ($g_{31}$), the results can be extended to any of the other connections by symmetry and interpreted accordingly without loss of generality. As seen previously with the symmetric motif, the rhythmic behaviors of this network also vary depending upon the gap between the nullclines, the transitions between soft and hard locking, as well as the release and escape mechanisms of the cells. Here and in subsequent discussions of other asymmetric motifs, we will show a bifurcation diagram varying the parameters $g$ (or $g_{31}$ in this case) and $I_{app}$, that effectively depicts a range of rhythmic behaviors exhibited by the network. We then pick a few parameter values to demonstrate detailed phase lag return maps and their transitions for cells obeying the release mechanism, as the synaptic strength is gradually increased (Fig.~\ref{fig:singrel}). This is then followed by a similar demonstration for cells obeying the escape mechanism. For these asymmetric network configurations, we observe asymmetric bifurcations in which only one or two pacemakers may appear or disappear, rather than all pacemakers (or all traveling waves) simultaneously, as seen in the fully symmetric network for any given parameters. 

Figure~\ref{fig:bifsingle} represents the bifurcation diagram for the mono-biased motif (shown in Fig.~\ref{fig:AsymMotifs}A), as the parameters $I_{app}$ and $g_{31}$ are varied, while the remaining synaptic connections are held constant at $g_{all}$=0.001. The network can produce several distinct multistable behaviors composed of just pacemakers (PM), just traveling waves (TW), just phase slipping (PS), or a combination of pacemakers with phase slipping (PM/PS) or traveling waves (PM/TW). Transitions between these regions are due to saddle-node (SN) bifurcations eliminating or restoring FPs to the map. The points $A-F$ highlighted along the dotted line near the top of the bifurcation diagram indicate the parameter values used for the phase lag return maps for the cells obeying the escape mechanism (Fig.~\ref{fig:singesc}), while the points A\textprime{}--F\textprime{} on the dashed line near the bottom highlight the parameter values for the cells obeying the release mechanism (Fig.~\ref{fig:singrel}). The vertical dotted line represents the parameter values where the network retains full symmetry, with $g_{31}$=$g_{\rm all}$. As such, the rhythmic behaviors and the transitions as we move along this line are identical with the fully symmetric motif. 

For cells obeying the release mechanism shown in Fig.~\ref{fig:singrel}, we initially disable the synapse $g_{31}$=0, while all other synaptic connection strengths are held constant at $g_{\rm all}$=0.001. In this case, we observe that the network is dominated by the green PM rhythm with the largest basin of attraction, while the blue PM rhythm is also stable although with a smaller basin (Fig.~\ref{fig:singrel}A). One may observe the presence of two saddle nodes (gray diamonds) and an interesting pattern of whorls in the phase space near the original location of the purple TW (0.66, 0.33), that is currently not seen in the network. Restoring the missing synapse $g_{31}=0.00081$ in Fig.~\ref{fig:singrel}B\textprime{} increases the blue PM basin, and also leads to the formation of the purple TW pattern through a saddle-node bifurcation, that gives rise to a third saddle around the purple TW FP. Strengthening of this synapse at $g_{31}=0.00108$ leads to the appearance of another saddle node and a fixed point corresponding to the red pacemaker rhythm, both of which rapidly diverge (see Fig.~\ref{fig:singrel}C\textprime{}). With further increases in the strength of $g_{31}$ through  ${0.00135, 0.004, 0.008}$ in Figs.~\ref{fig:singrel}D\textprime{}--F\textprime{}, the purple TW becomes the dominant rhythm of the network via a series of saddle-node bifurcations where the red, green and blue PM FPs, respectively, merge with the three saddles surrounding the purple TW and disappear one after the other.

Figure~\ref{fig:singesc} shows the evolution of Poincar\'{e} return maps for cells obeying the escape mechanism as $g_{31}$ gradually increases. In Fig.~\ref{fig:singesc}$A$ with $g_{31}$=0, the network is dominated by a single clockwise TW (black FP). Fig.~\ref{fig:Black4Panels} shows 4 identical panels (same as Fig.~\ref{fig:singesc}$A$) stitched together to aid the visual inspection of the trajectories and their convergence. A saddle (gray diamond) exists very close to the black TW FP in the phase space, which gives rise to interesting dynamics such that two trajectories (red lines) starting from very close initial conditions (on either side of the blue line), take entirely different paths but ultimately converge to the same fixed point (black TW). Figure~\ref{fig:singEscPhaseSlip4Panels} shows a similar alignment of four identical return maps for this network, but at a slightly higher value of $I_{app}=0.5875$, chosen within the phase slipping (PS) region of the bifurcation diagram in Fig.~\ref{fig:bifsingle}, close to the parameter values corresponding to the map depicted in Fig.~\ref{fig:singesc}A. Comparing this with Fig.~\ref{fig:Black4Panels} shows that the black TW FP and the saddle merge and disappear following a saddle-node bifurcation, and give rise to a stable PS pattern in Fig.~\ref{fig:singEscPhaseSlip4Panels}. The network has no phase-locked rhythms in this state and all the trajectories from different initial conditions converge on to the red stable PS invariant circle, after an initial transient. Conversely, the blue line represents an unstable PS pattern. If a trajectory starts with initial conditions exactly on this blue line, it will continue to move along this unstable PS pattern, but slight perturbations would lead to diverging paths that then ultimately converge on to the stable red PS pattern. Note that within these PS patterns, cells 2 and 3 remain nearly phase-locked, while cell 1 undergoes phase-slipping and so runs at a different frequency compared to the other two cells. Due to the existence of such subpopulations of cells that run at distinct frequencies, PS has also been referred to as a chimera state. Stable and unstable PS patterns are elaborated further in the following sections.

As the coupling strength $g_{31}$ increases from $0$ to $0.00027$ in Fig.~\ref{fig:singesc}B, the unstable PS pattern first undergoes a saddle-node bifurcation, giving rise to an unstable fixed point (white dot) and a saddle. The saddle then undergoes pitch-fork bifurcation to give rise to the blue pacemaker rhythm, as well as two additional saddles. Further increases in $g_{31}$ give rise the red and green pacemakers, following their respective saddle-node bifurcations in Fig.~\ref{fig:singesc}C and Fig.~\ref{fig:singesc}D. The basin of attraction of the black TW continues to diminish and undergoes an additional torus bifurcation in Fig.~\ref{fig:singesc}E, creating an invariant circle, while the green PM rhythm is lost following a saddle-node bifurcation, resulting in a larger red PM basin. In Fig.~\ref{fig:singesc}F, the invariant circle and the black TW rhythm are finally lost as the black repellor converges with a saddle, and the system is dominated by just the red and blue PMs, with very rapid convergence for many initial conditions (exemplified by large white regions on the return map, where convergence is so rapid that traces are not apparent in these areas).

\subsection{Double-biased motif}

In this section, we study the dynamics of the double-biased motif (Fig.~\ref{fig:AsymMotifs}B), where asymmetry in the network is achieved by simultaneously altering a pair of connections, $g_{31}$ and $g_{13}$ between cell 1 and 3.  Fig.~\ref{fig:bifdouble} shows the bifurcation diagram for this motif, with the sampled parameter values for the release (Fig.~\ref{fig:doublerel}) and the escape (Fig.~\ref{fig:doubleesc}) mechanisms highlighted along the bottom dashed and top dotted lines, respectively. The vertical dotted line represents symmetric network configuration as described previously. The bifurcation diagram reveals that the network can produce just pacemakers (PM), just traveling waves (TW), just phase slipping (PS), or any combination of a pair of such rhythms (PM/PS, PM/TW, TW/PS). Pacemaker behavior dominates at weak coupling, while phase slipping does so at strong coupling. Other rhythms exist primarily near the mid-ranges of values for $I_{app}$, or close to full symmetry for the values of $g$. 

\begin{figure}[h!]
\begin{center}
\resizebox*{0.999\columnwidth}{!}
{\includegraphics{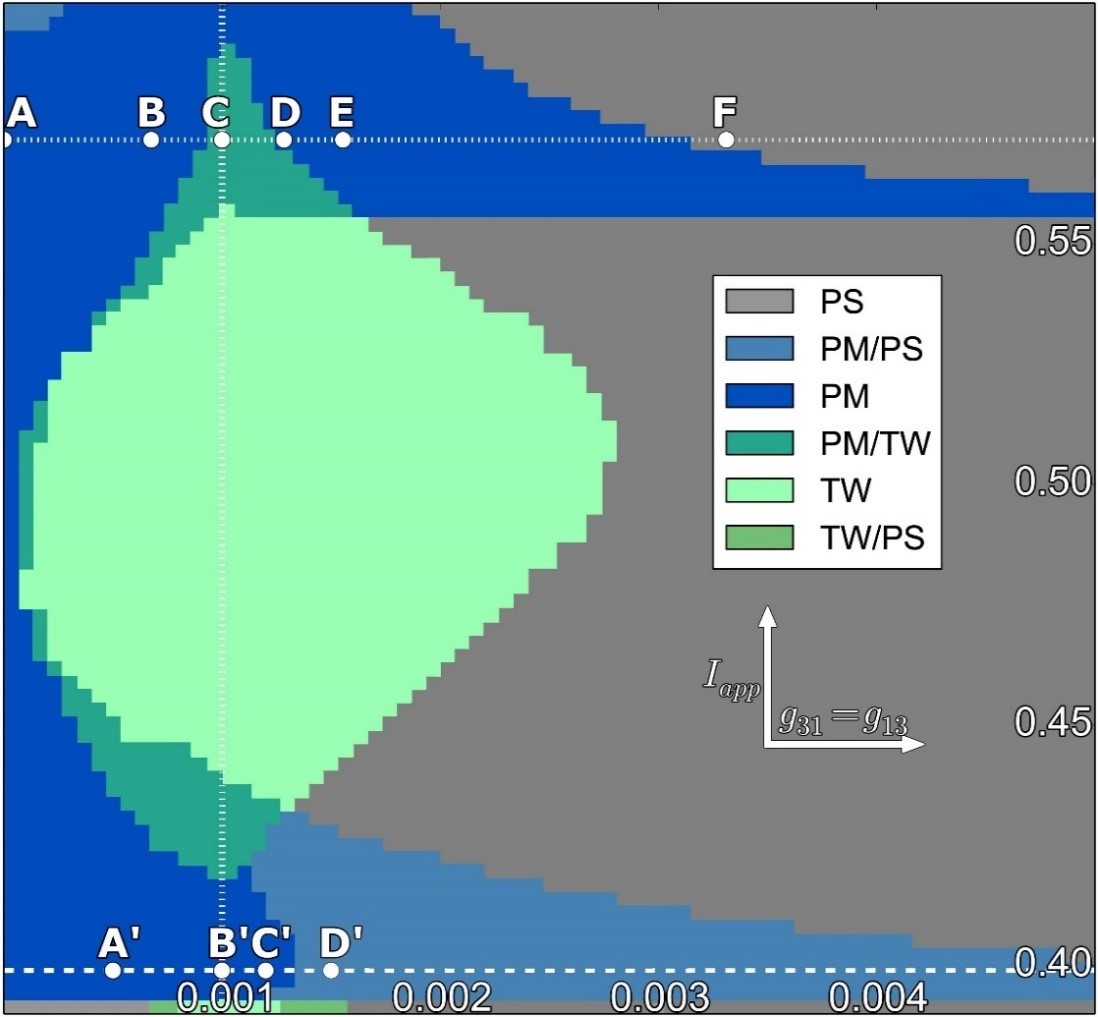}}
\caption{The $(g_{31/13}, \, I_{app})$-bifurcation diagram of the double-biased motif. The network can produce just pacemakers (PM), just traveling waves (TW), just phase slipping (PS), or any combination of a pair of such rhythms: PM/PS, PM/TW, and TW/PS in the color-mapped regions. The PM and PS behaviors dominate at weak and strong coupling, respectively. Points A\textprime{}--D\textprime{} and A-F indicate the sampled parameter values for the release (Fig.~\ref{fig:doublerel}) and the escape (Fig.~\ref{fig:doubleesc}) mechanisms, respectively. The vertical line is where  the network is symmetric at $g_{31}$=$g_{\rm all}=0.001$.}
\label{fig:bifdouble}
\end{center}
\end{figure}

\begin{figure}[h!]
\begin{center}
\resizebox*{1.\columnwidth}{!}{\includegraphics{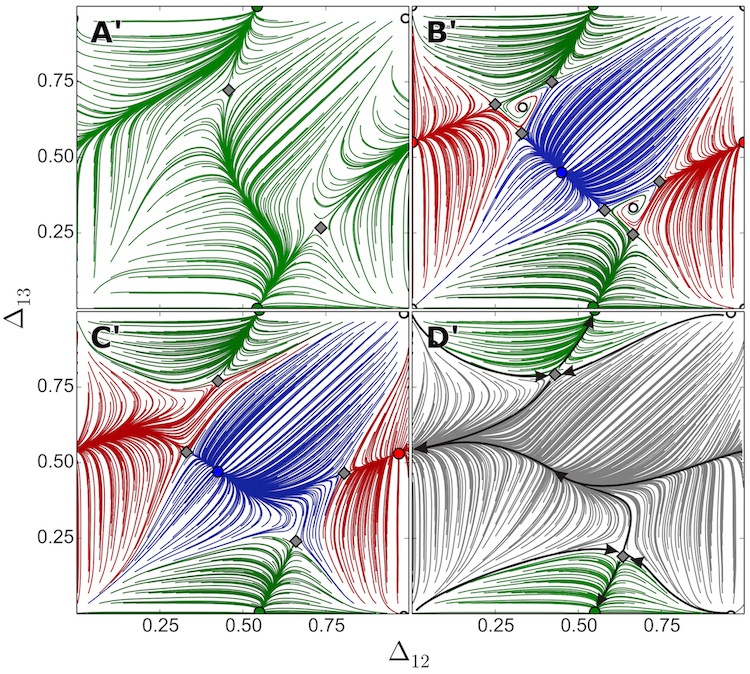}}
\caption{Poincar\'{e} return maps corresponding to the double-biased motif with the release mechanism. When the synapses $g_{31}$ and $g_{13}$ are weak, there is only the green PM in the map shown in Panel~A\textprime{}. As their strength is increased, the blue and red PMs emerge through a series of saddle-node and pitch-fork bifurcations, along with multiple saddles and two repelling FPs corresponding to the unstable TWs in Panel~B\textprime{}. Further increase in the synaptic strengths causes the unstable TW FPs to disappear through a heteroclinic saddle-node bifurcation in Panel~C\textprime{}. The blue and red PMs also disappear through a heteroclinic saddle-node bifurcation, thus giving rise to a stable ``phase-slipping'' invariant circle (shown gray), coexisting with the green PM in Panel D\textprime{}. Their basins are partitioned by  the incoming separatrices (black curves) of the saddles. Parameters: $I_{app} = 0.399$, $g_{\rm all} = 0.001$ except for $g_{31} = g_{13}$= (0.0005, 0.001, 0.0012, 0.0015). 
}
\label{fig:doublerel}
\end{center}
\end{figure}

\begin{figure}[h!]
\begin{center}
\resizebox*{1.0\columnwidth}{!}{\includegraphics{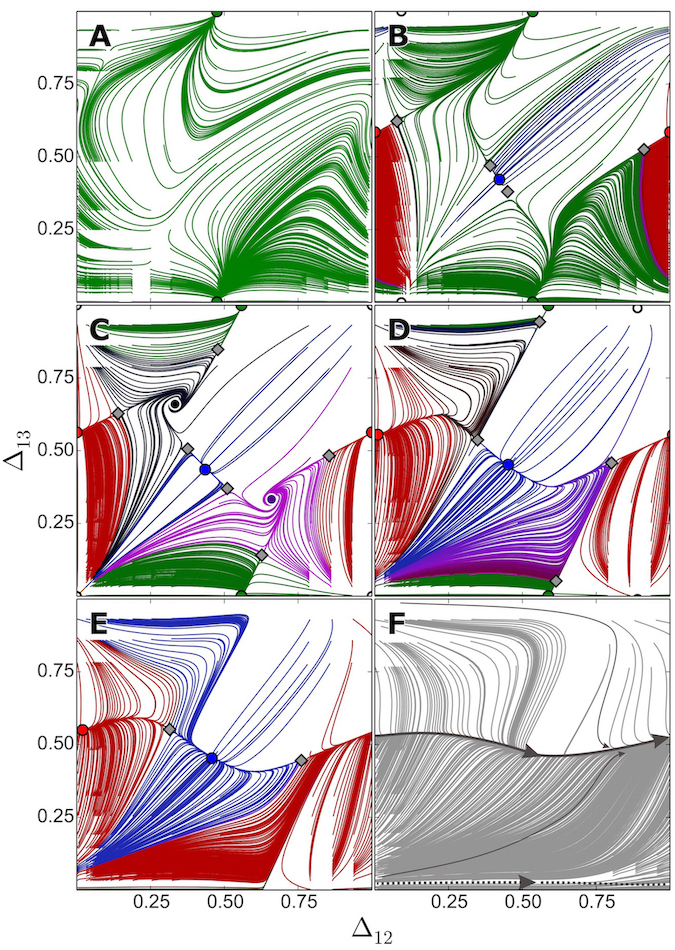}}
\caption{Poincar\'{e} return maps corresponding to the double-biased motif with the escape mechanism. When both synapses $g_{31} = g_{13}=0$, the network produces only the green PM rhythm in panel~(A) (elaborated further in Fig.~\ref{fig:Green4Panels}). The blue and red PMs emerge following pitch-fork bifurcations of the saddles in panel B. The two TWs emerging in panel~C,  disappear through saddle-node bifurcations in panel~D. Next, the green PM disappears via a saddle-node bifurcation, after merging with a nearby saddle in panel~E. The red and blue PMs finally disappear through a heteroclinic saddle-node bifurcation, that gives rise to the only stable invariant circle  (gray) in panel~F, moving in the opposite direction compared with Fig.~\ref{fig:doublerel}D\textprime{}. Parameters: $I_{app} = 0.5716$, $g_{\rm all} = 0.001$, except for $g_{31} = g_{13}$= (0, 0.000676, 0.001, 0.00128, 0.00155, 0.00331). }
\label{fig:doubleesc}
\end{center}
\end{figure}

\begin{figure}[h!]
\begin{center}
\resizebox*{0.99\columnwidth}{!}
{\includegraphics{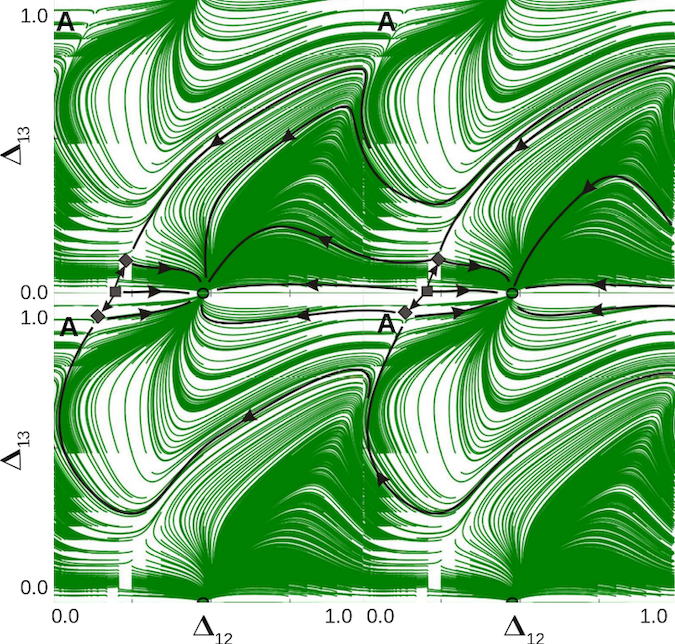}}
\caption{Four identicals panels from Fig.~\ref{fig:doubleesc}A stitched together to continuously visualize trajectories wrapping around the 2D torus. This network is mono stable with a single PM rhythm -- the green FP at (0.5,0), to which some trajectories converge along quite a long path, as its attraction basin is shaped by complex interactions of the separatrices (black lines) of the two saddles (black $\diamond$).}
\label{fig:Green4Panels}
\end{center}
\end{figure}

Figure~\ref{fig:doublerel} shows the evolution of the Poincar\'{e} return maps for the release mechanism as the synaptic strengths of both $g_{31}$ and $g_{13}$ increase from $0.0$ to $0.0045$, while all other synaptic strengths remain constant at $0.001$. When both the synapses are absent or very weak ($g_{31}=g_{13}=0.0005$) in Fig.~\ref{fig:doublerel}A\textprime{}, the network produces a single stable rhythm of the green PM. This could be inferred from the fact that only cell 2 (green) has outgoing inhibitory connections to the other two cells. As we strengthen the coupling ($g_{31}=g_{13}=0.001$) in Fig.~\ref{fig:doublerel}B\textprime{}, following a series of saddle-node and pitch-fork bifurcations, the blue and red PMs emerge along with multiple saddles  and two repelling FPs corresponding to unstable TWs. Further increase in the synaptic strength ($g_{31}=g_{13}=0.0012$) in Fig.~\ref{fig:doublerel}C\textprime{} causes the unstable TW FPs to disappear through saddle-node bifurcations and the corresponding increase in the blue and red PM basins. In Fig.~\ref{fig:doublerel}D\textprime{} with $g_{31}=g_{13}=0.0015$, the blue and red PMs disappear and give rise to a stable invariant circle (gray) through a heteroclinic saddle-node bifurcation. This invariant circle corresponds to a PS rhythmic pattern that wraps around the torus, and coexists along with the green PM. Hand-drawn lines in black are sampled to illustrate the attraction basins bounded by the incoming separatrices of the saddles. 

Figure~\ref{fig:doubleesc} illustrates the dynamics and the corresponding return maps for the escape mechanism as the synaptic strengths $g_{31}$ and $g_{13}$ are gradually increased through (0, 0.000676, 0.001, 0.00128, 0.00155, 0.00331) at the points from A through F. When both the synapses are turned off in Fig.~\ref{fig:doubleesc}A, the network produces a single stable rhythm with the green PM, as described previously. Figure~\ref{fig:Green4Panels} shows 4 identical panels (same as in Fig.~\ref{fig:doubleesc}A, with greater detail) placed next to each other to aid visual inspection of the trajectories and their convergence. Sample trajectories are shown in black, along with the green pacemaker, two saddles, and an unstable fixed point. In Fig.~\ref{fig:doubleesc}B, the two saddles undergo pitch-fork bifurcations to give rise to the blue and red PMs, as well additional saddles. As the synapses are further strengthened in Fig.~\ref{fig:doubleesc}C--D, the two TWs emerge and then disappear through saddle-node bifurcations. Further increases lead gradually to the disappearance of the green pacemaker in Fig.~\ref{fig:doubleesc}E via a saddle-node bifurcation, and of the red and blue PMs in Fig.~\ref{fig:doubleesc}F through a  heteroclinic saddle-node bifurcation that gives rise to a single invariant circle of the gray phase-slipping pattern that wraps around the torus.

\subsection{Driver-biased motif}

\begin{figure}[h!]
\begin{center}
\resizebox*{0.99\columnwidth}{!}
{\includegraphics{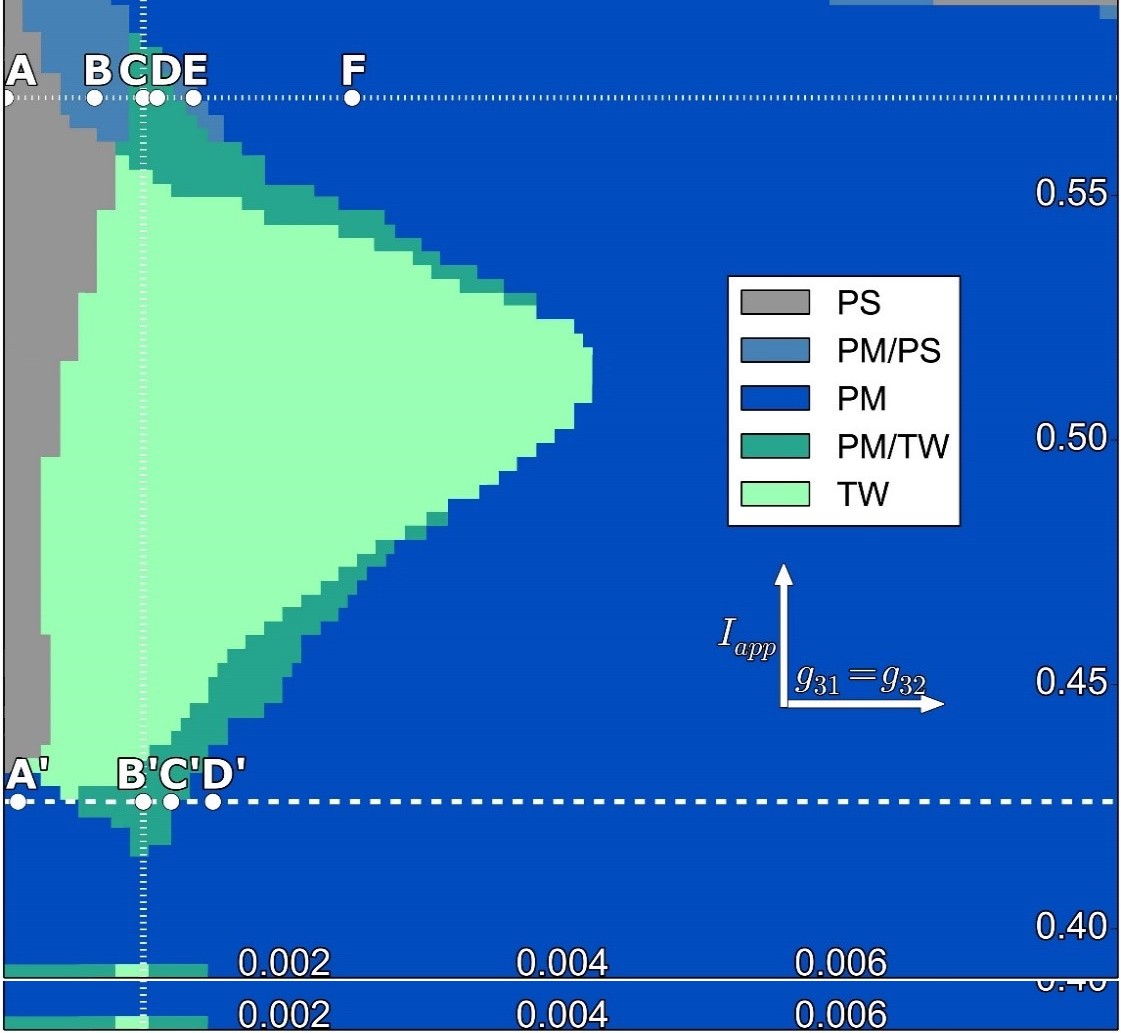}}
\caption{$(g_{31/32},\,I_{app})$-bifurcation diagram for the driver-biased motif. The network is dominated by the red PM for strong synaptic coupling, as expected from the asymmetry. The network also produces TWs, PS, and a combination of PMs with TWs or PS at weaker coupling, along with the blue and green PMs. The vertical line represents the network symmetry where $g_{31} = g_{32} = g_{\rm all} = 0.001$. Sampled parameter values for the release and escape mechanism are as described previously and elaborated in Fig.~\ref{fig:kingrel} and Fig.~\ref{fig:kingesc}, respectively.}
\label{fig:bifking}
\end{center}
\end{figure}

\begin{figure}[h!]
\begin{center}
\resizebox*{0.99\columnwidth}{!}
{\includegraphics{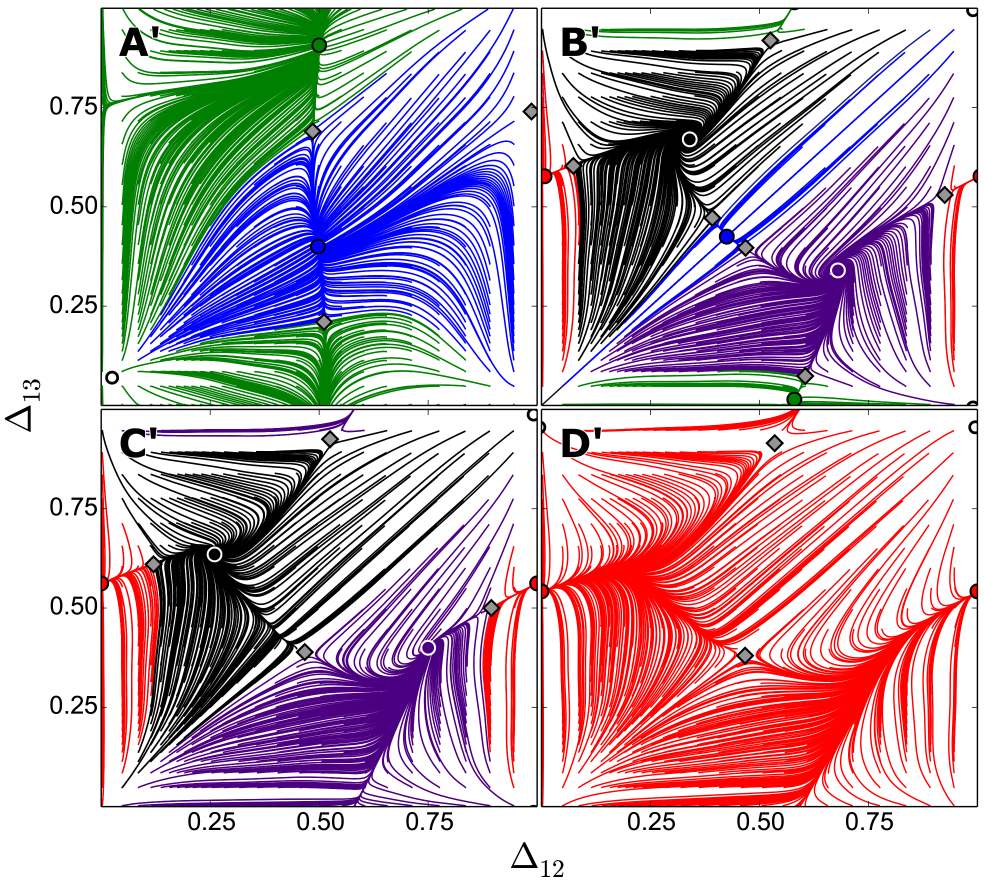}}
\caption{Poincar\'{e} return maps corresponding to the driver-biased motif with the release mechanism. For small $g_{31}=g_{32}$-values,  the network produces both the blue and green PMs in panel A\textprime{}. As their value is increased, the red PM and the two TWs then emerge via a series of saddle-node bifurcations in panel~B\textprime{}. Next, the green and the blue PMs disappear in C\textprime{},  as well as the black and purple traveling waves in D\textprime{}, via saddle-node bifurcations, resulting in the single dominant red PM. Parameters: $I_{app} = 0.426$, $g_{\rm all} = 0.001$ except for $g_{31} = g_{32}$=(0.0001, 0.001, 0.00115, 0.0015).}
\label{fig:kingrel}
\end{center}
\end{figure}

\begin{figure}[h!]
\begin{center}
\resizebox*{0.99\columnwidth}{!}
{\includegraphics{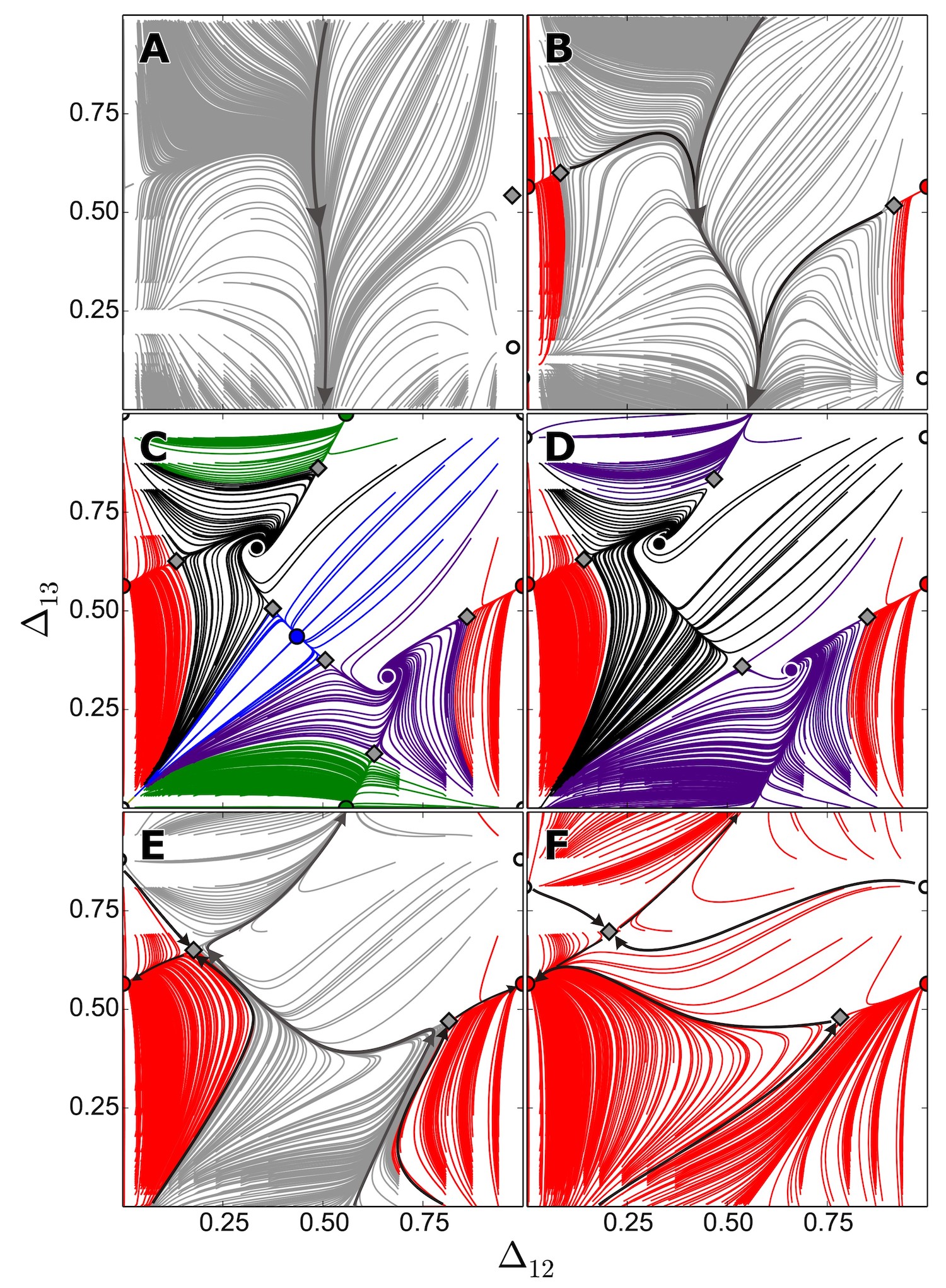}}
\caption{Poincar\'{e} return maps corresponding to the driver-biased motif with the escape mechanism: with weak synapses $g_{31} = g_{32}$, the network is dominated by a single phase-slipping rhythm in panel A. As they become stronger, through a series of saddle-node bifurcations, the red PM first emerges in (B), followed by the blue and green PMs, and the two TWs in (C). Further, the blue and the green PMs then disappear in (D), followed by the loss of the two TWs in (E), via a heteroclinic saddle-node bifurcation, giving rise to a stable PS pattern -- the stable IC (grey basin) that coexists with the red PM. The IC then disappears and the network becomes mono-stable with the  single dominant PM rhythm corresponding to the red FP in (F). Parameters: $I_{app} = 0.57$, $g_{\rm all} = 0.001$ except for $g_{31}=g_{32}$=(0.00001, 0.00065, 0.001, 0.0011, 0.00136, 0.0025).}
\label{fig:kingesc}
\end{center}
\end{figure}

Another type of asymmetry we investigate is the driver-biased motif, where the two outgoing synapses from cell 3 ($g_{31} = g_{32}$) are manipulated, while the remaining connection strengths are held constant. Figure~\ref{fig:bifking} shows the bifurcation diagram and the sampled values in it, corresponding to the release and escape mechanisms, as described previously. As can be expected from this asymmetry, for sufficiently strong synaptic coupling for the outgoing connections from cell 3 (red) which acts as the driver, the network is dominated by the red PM rhythm. For weaker coupling strengths, one can observe TWs, PS, and a combination of PMs with TWs or PS. At weaker coupling, one may also see the blue and green pacemaker rhythms.

Figure~\ref{fig:kingrel} shows the return maps for the cells obeying the release mechanism. For small values of $g_{31}=g_{32}$ in Fig.~\ref{fig:kingrel}A\textprime{}, the blue and green PMs coexist. As the synaptic strengths are increased in Fig.~\ref{fig:kingrel}B\textprime{}, the red PM and the two TWs emerge following a series of saddle-node bifurcations. Further increases in the synaptic strengths lead to the disappearance  of the original blue and green PMs in Fig.~\ref{fig:kingrel}C\textprime{}, as well as the disappearance of the two TWs in Fig.~\ref{fig:kingrel}D\textprime{} through saddle-node bifurcations, resulting in a single red PM rhythm of the circuit, under the control of the biased-driver cell.

Following the return maps for the escape mechanism in Fig.~\ref{fig:kingesc}A--F as the values of $g_{31}$ and $g_{32}$ are simultaneously increased, shows that the network is initially dominated by a single PS rhythm in Fig.~\ref{fig:kingesc}A. A red PM rhythm then emerges following a saddle-node bifurcation in Fig.~\ref{fig:kingesc}B. A series of saddle-node bifurcations then gives rise to the blue and green PMs, as well as the two TWs in Fig.~\ref{fig:kingesc}C. With further increase in synaptic strengths, the blue and green PMs first disappear in Fig.~\ref{fig:kingesc}D, followed by the loss of the two TWs in Fig.~\ref{fig:kingesc}E, through saddle-node bifurcations, resulting in a phase slipping pattern that coexists with the red PM. Finally, for very strong coupling in Fig.~\ref{fig:kingesc}F, the phase slipping pattern also disappears, giving rise to a single red PM rhythm for the circuit, driven by the dominant cell.

\subsection{Clockwise-biased motif}
\begin{figure}[h!]
\begin{center}
\resizebox*{0.99\columnwidth}{!}
{\includegraphics{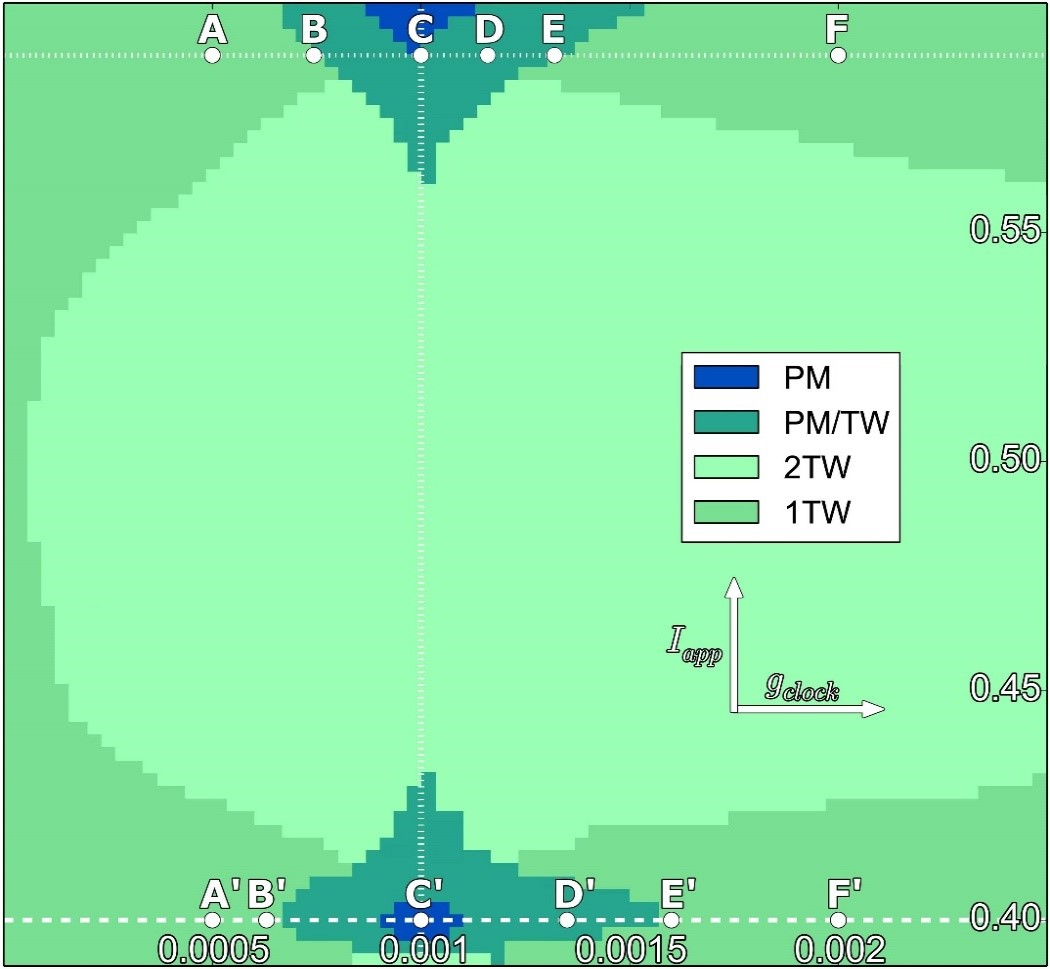}}
\caption{$(g_\circlearrowright , \, I_{app}$-bifurcation diagram of the clockwise-biased motif. The network is dominated primarily by traveling wave rhythms as expected from this asymmetry. A single TW is seen for both small and large $g_\circlearrowright$-values, while both the TWs are seen for moderate values. PMs and PM/TWs are also seen close to symmetry in the network, indicated by the vertical dotted line where $g_\circlearrowright=g_{\rm all}=0.001$. Parameter values sampled for the release and the escape mechanisms in Fig.~\ref{fig:clockrel} and Fig.~\ref{fig:clockesc} are also shown.}
\label{fig:bifclock}
\end{center}
\end{figure}

\begin{figure}[h!]
\begin{center}
\resizebox*{0.99\columnwidth}{!}
{\includegraphics{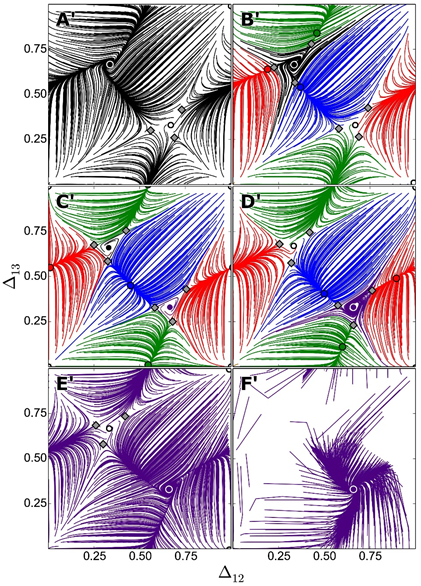}}
\caption{Poincar\'{e} return maps corresponding to the clockwise-biased motif with release mechanism. The network is initially dominated by a single stable clockwise TW in panel A\textprime{}, while the purple TW FP remains unstable. As the clockwise synapses are strengthened, the three PMs emerge in panel B\textprime{} via saddle-node bifurcations around the black TW FP. The purple TW then becomes stable in (C\textprime{}), while the black TW  in (D\textprime{}) loses stability, via torus bifurcations. The three PMs disappear via saddle-node bifurcations in panel~E\textprime, resulting in the single dominant purple TW rhythm that finally becomes hard-locked in (F\textprime{}) as indicated by the rapid, non-smooth convergence of trajectories to it. Parameters: $I_{app} = 0.4$, $g_{\rm all} = 0.001$ except $g_{12} = g_{23} = g_{31}$= (0.0005, 0.00065, 0.001, 0.00135, 0.0016, 0.002).}
\label{fig:clockrel}
\end{center}
\end{figure}

\begin{figure}[h!]
\begin{center}
\resizebox*{0.99\columnwidth}{!}
{\includegraphics{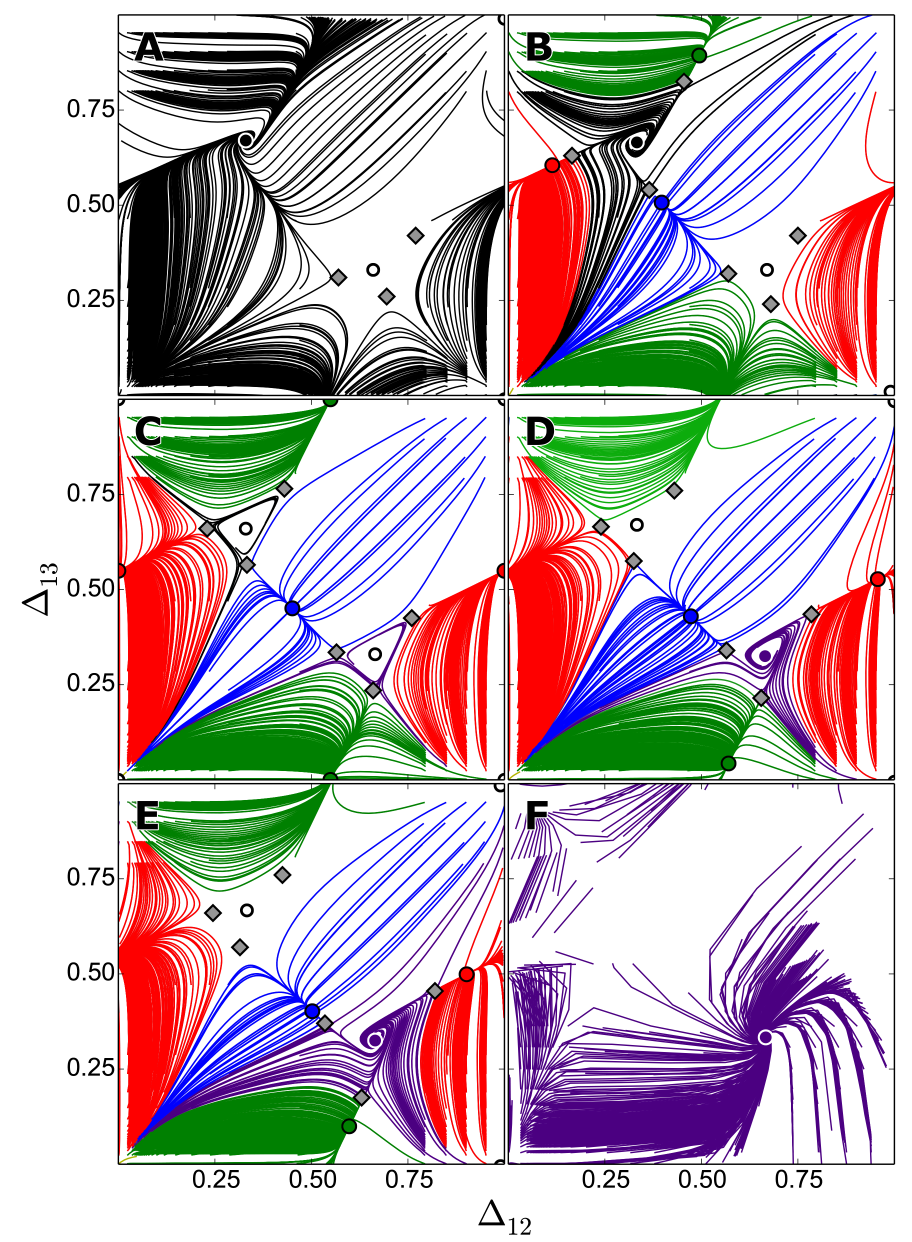}}
\caption{Poincar\'{e} return maps corresponding to the clockwise-biased motif with the escape mechanism look almost identical to those with the release mechanism shown in Fig.~\ref{fig:clockrel}. The network transitions from a single dominant black TW in panel~A to purple TW in panel~F. Along the way, the three PMs emerge via saddle-node bifurcations in (B), the black TW loses stability in (C), while the purple TW becomes stable in panels~D and E via torus bifurcations (resulting in the transitory invariant circle shown), and the three PMs are lost via saddle-node bifurcations $(F)$. Parameters: $I_{app} = 0.5886$, $g_{\rm all} = 0.001$ except $g_{12} = g_{23} = g_{31}$= (0.0005, 0.000743, 0.001, 0.00116, 0.00132, 0.002).}
\label{fig:clockesc}
\end{center}
\end{figure}

The next asymmetric network configuration examined is the clockwise-biased motif, in which all the clockwise connections: $g_{12}$, $g_{23}$, and $g_{31}$ are manipulated simultaneously, while the anti-clockwise connections are held constant. Figure~\ref{fig:bifclock} shows the bifurcation diagram for this motif, which reveals that the TW rhythms dominate this network. A single TW is seen at either end of the coupling strength spectrum, as can be expected from such asymmetry, while both the TWs are seen in between these two regions. PMs and PM/TW combinations are also seen in parametric regions close to the fully symmetric network configuration, for both the escape and release mechanisms. 

Fig.~\ref{fig:clockrel} shows the return maps at parameter values sampled for the release mechanism, as the clockwise synapses are gradually strengthened. Initially, the network is dominated by a single clockwise TW (black) in Fig.~\ref{fig:clockrel}A\textprime{}. Following a series of saddle-node bifurcations, the three PMs emerge in Fig.~\ref{fig:clockrel}B\textprime{}. This is followed by the emergence of the anti-clockwise TW (purple) in Fig.~\ref{fig:clockrel}C\textprime{} and the disappearance of the clockwise TW (black) in Fig.~\ref{fig:clockrel}D\textprime{}, through their respective torus bifurcations. The three PMs then disappear via saddle-node bifurcations in Fig.~\ref{fig:clockrel}E\textprime{}, leading to a single anti-clockwise TW rhythm dominating the network. Further strengthening of the synapses in Fig.~\ref{fig:clockrel}F\textprime{} leads to hard locking, with a single TW rhythm.

Fig.~\ref{fig:clockesc} depicts the dynamical transitions for the escape mechanism, and looks almost identical to that of the release mechanism shown in Fig.~\ref{fig:clockrel}. The transition from the black clockwise TW in Fig.~\ref{fig:clockesc}A to the purple counter-clockwise TW in Fig.~\ref{fig:clockesc}F occurs via the formation of the three PMs, the disappearance of the black TW, the appearance of the purple TW, followed by the disappearance of the three PMs. The escape mechanism is more conducive to permitting the transitory limit cycle behavior following torus bifurcations and so these can be observed in finer detail here in Fig.~\ref{fig:clockesc}C.

\subsection{Emergence of a transient torus}

\begin{figure}[ht!]
\begin{center}
\resizebox*{0.98\columnwidth}{!}{\includegraphics{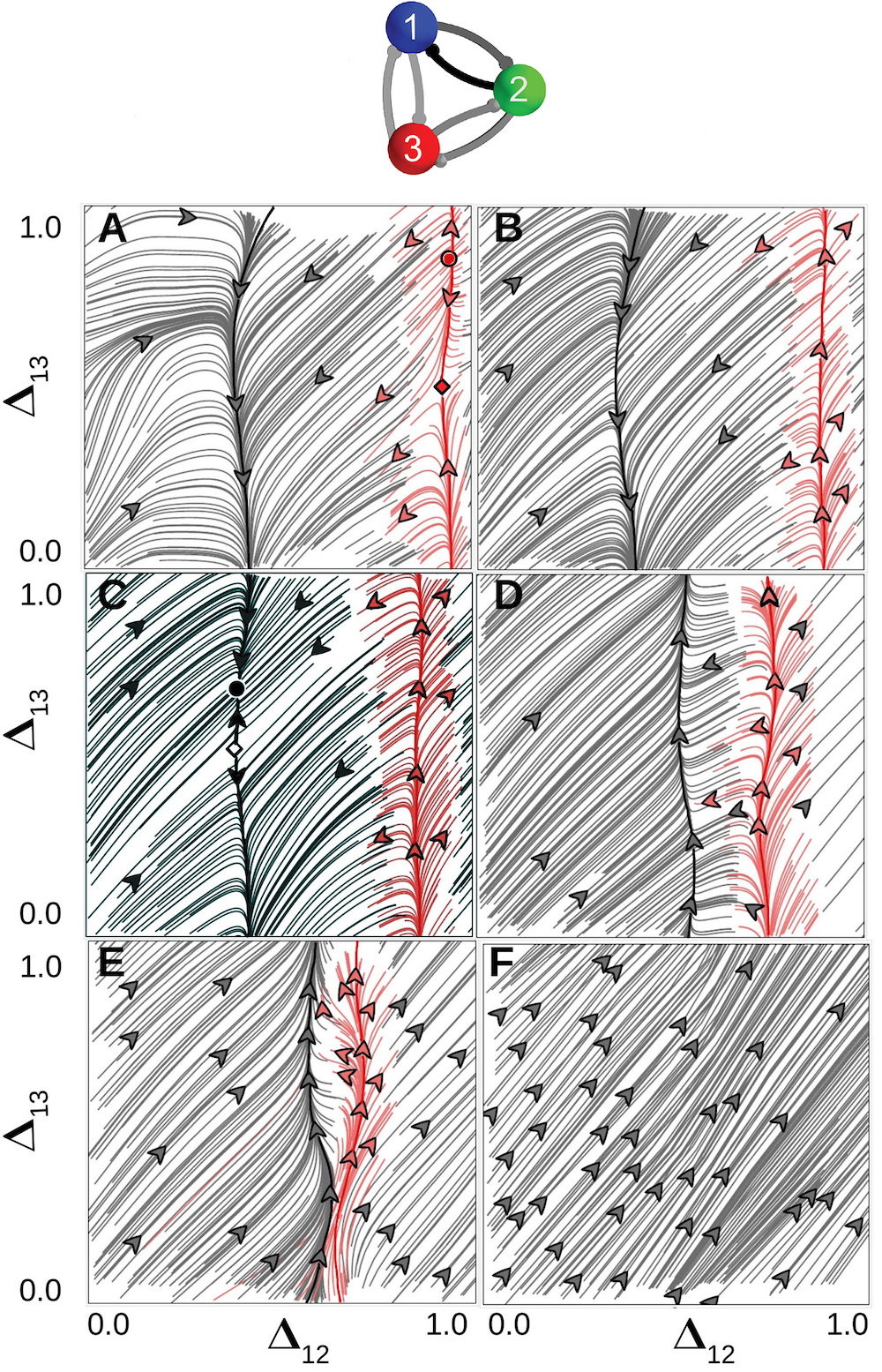}}
\caption{Six maps demonstrating the stages in the emergence of a transitive torus in the given asymmetric  motif (top). By increasing $I_{app}$, the network migrates from the release to the escape mechanisms. Both forward (gray) and backward (red) trajectories elucidate the unstable FPs and ICs. Step 1: The network initially has a single stable  phase-slipping pattern (vertical gray IC with the decreasing $\Delta_{13}$-direction) in panel~A, along with a repelling FP near the origin and a saddle. Step 2: both merge and vanish through a homoclinic saddle-node bifurcation to give rise to an additional repelling  PS pattern (red IC with the increasing $\Delta_{13}$-direction) in panel~B. Step 3: the stable IC breaks down via a  homoclinic saddle-node bifurcation as two FPs, a stable one and a saddle, emerge on it so that the outgoing separatrices of the saddle end up on the stable FP around (0.3, \, 0.6). The gap between both FPs increases, which makes them closer to each other after crossing the boundary $\Delta_{13}=0=1$. Step~4: The FPs disappear via a reverse homoclinic saddle-node bifurcation and the stable PS pattern re-emerges with the opposite direction, matching the orientation of the unstable IC in panel~D. Step~5: the distance between the stable and the unstable ICs decreases, as they start to merge and vanish in panel~E. This completes the bifurcation sequence giving rise to a transitive torus in (F), without any FPs or ICs such that a single trajectory densely fills its surface. Parameters: $g_{13}=g_{31}=0.0003$, $g_{23}=g_{32}=0.0005$, $g_{12}=001$, $g_{21}=0.0051$, $I_{app}$= (0.4,\, 0.46,\, 0.5,\, 0.572982,\, 0.594,\, 0.61).} 
\label{fig:ErgodicTorus}
\end{center}
\end{figure}

In this section, we describe a route to the emergence of a transient torus without any fixed points or invariant circles in the return map. For the asymmetric network under consideration (Fig.~\ref{fig:ErgodicTorus} top), this would correspond to a lack of any phase-locked or periodically varying rhythmic outcomes. Such a transient torus has what is the called {\it everywhere dense covering}, and a trajectory starting from any initial condition fills in the entire phase portrait over time.

Figure~\ref{fig:ErgodicTorus} (top) shows the asymmetric network, where cells 1 and 3 have weak inhibitory coupling ($g_{13}=g_{31}=0.0003$), cells 2 and 3 have a slightly stronger coupling ($g_{23}=g_{32}=0.0005$), while cells 1 and 2 have stronger asymmetric coupling ($g_{12}=0.001$, and $g_{21}=0.0051$). Unlike the previous examples, we maintain the synaptic strengths constant in this case, while varying the external current drive $I_{app}$ for all the cells simultaneously through ${0.4, 0.46, 0.5, 0.572982, 0.594, 0.61}$ from panel A through panel F in Fig.~\ref{fig:ErgodicTorus}, transforming individual cells gradually from the release to the escape mechanisms. In order to effectively demonstrate both the stable and unstable fixed points, invariant circles, as well as their transitions, we compute both the forward (gray) and backward (red) trajectories and plot them in the Poincar\'{e} return maps. 

In Fig.~\ref{fig:ErgodicTorus}A, the network produces a single stable invariant circle or a phase slipping pattern, shown in gray. In addition, the red trajectories show the presence of an unstable fixed  point as well as a saddle, both of which disappear after undergoing a saddle-node bifurcation in Fig.~\ref{fig:ErgodicTorus}B to give rise to an unstable PS pattern, that coexists with the stable PS. Also, note that the stable and unstable PS patterns run in opposite orientations. In Fig.~\ref{fig:ErgodicTorus}C, the stable PS pattern is lost due to a saddle-node bifurcation, giving rise to a stable fixed point and a saddle. The network therefore has a single stable rhythm (the black TW rhythm, shown in previous examples) that coexists with an unstable PS pattern. The fixed point and the saddle then gradually move further apart from each other, wrap around the torus, and then merge and disappear through a saddle-node bifurcation in Fig.~\ref{fig:ErgodicTorus}D, again giving rise to a stable PS pattern, that runs along the same orientation as the unstable PS pattern. The PS patterns then starts disappearing via torus breakdown \cite{ju2018bottom} as the stable and unstable invariant circles start merging in Fig.~\ref{fig:ErgodicTorus}E, and finally give rise to the transient torus in 
Fig.~\ref{fig:ErgodicTorus}F. Only a single long trajectory is plotted in Fig.~\ref{fig:ErgodicTorus}F and it retraces the entire phase space over time, without any fixed points or invariant circles.

\subsection{Invisible heteroclinic bifurcations}

Here, we illustrate the ``invisible'' role of some heteroclinic bifurcations of the saddles and how they determine the structure of the 2D Poincare return maps, and specifically how they shape the attraction basins of the co-existing stable FPs. Let us examine the transformation of the attraction basins depicted in Fig.~\ref{fig:switch}. In both the cases, the map has two stable FPs, green and blue, corresponding to the pacemaker rhythms with phase lags  $ \sim (0.5,\, 0)$ and $ \sim (0.5, \, 0.5)$, respectively. In addition to the persistent repeller at the origin, there are three more saddle FPs (labelled by $\diamond$'s) so that the total number of hyperbolic FPs on the torus is even. Of special interest here is the saddle to the left of the stable (blue) FP around $(0.5, \, 0.5)$. More specifically, let us follow its left outgoing (unstable) separatrix (set) to find its destination, or $\omega$-limit set, as the number of iterates increases. As the map on the torus is defined on mod~1, the separatrix disappears when it reaches the left wall given by $\Delta_{12}=0$, and comes back into the map from the right wall given $\Delta_{12}=1$. Next, it slides above the incoming separatrix of the other saddle (to the right of the blue FP) to converge to the blue FP -- its $\omega$-limit set (Fig.~\ref{fig:switch}A). As $g_{31}$ is slightly increased from 0.0071 to 0.0072, this separatrix first merges with the incoming separatrix of the other saddle (right) to form a one-way heteroclinic connection, which is followed by its shift further downwards below the saddle to switch to another $\omega$-limit set -- the stable FP around $ (0.5,\, 0)$, which corresponds to the green PM (Fig.~\ref{fig:switch}B). This heteroclinic bifurcation drastically repartitions the sizes of the attraction basins of the co-existing FPs: the blue FP is no longer dominant as a majority of initial conditions would now converge to the green FP. This example highlights the pivotal role of homoclinic and heteroclinic bifurcations that underlie major reconstructions of the phase space, while preserving the existence and stability of FPs in these return maps and other systems, in general. 

\begin{figure}[ht]
\begin{center}
\resizebox*{0.99\columnwidth}{!}
{\includegraphics{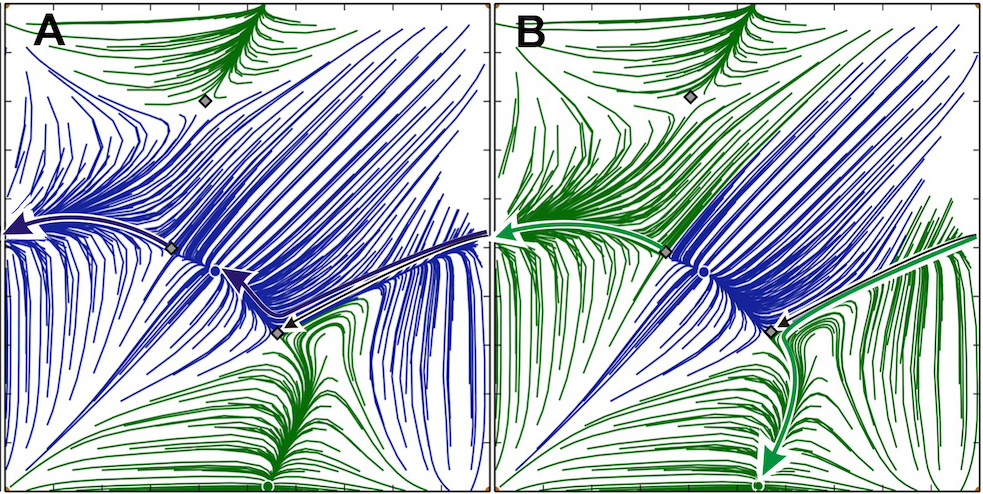}}
\caption{2D return maps showing how the outgoing and the incoming separatrices (grey lines wrapping around the torus) of the two saddles (labelled with grey $\diamond$)  shape the boundaries and the sizes of the attraction basins of the stable FPs, green and blue.  The direction in which the heteroclinic connection between the two saddle FPs splits, determines which stable FP has the largest attraction basin: blue in panel A at $g_{31}=0.0071$ or green in panel B at $g_{31}=0.0072$; other parameters: $g_{12}=g_{32}=0.0038$ and $g_{21}=g_{23}=g_{13}=0.0041$, and $\epsilon=0.3$. }
\label{fig:switch}
\end{center}
\end{figure}

\subsection{Stable synchronous state}
	In this section, we discuss the unexpected case of a stable synchronous state, where all the cells oscillate together, that has not been previously observed in 3-cell circuits coupled by the {\em fast} FTM synapses. We note however that the discovery of the stable synchronous state, in addition to several other exponentially stable polyrhythms, was recently reported in 4-cell inhibitory circuits, made up of gFN-model neurons \cite{pusuluri2019computational}. So far, we have known that, as soon the fast inhibitory connections in 3-cell motifs are replaced with excitatory ones \cite{wojcik2014key}, all the attractors of the map become repellers, the incoming (stable) separatrices (sets) of the saddles become outgoing (unstable), and vice versa. This is also true for half-center oscillators made up of two reciprocally inhibitory cells. It was shown previously \cite{rubin2000geometric} that the synchronous state in such a HCO made up of two plain FN-oscillators is unstable, in the case of FTM or similar inhibitory coupling. Note that we say a synapse is fast when the corresponding current is only slightly delayed compared to the timing of the spike or the burst that initiates it, and decays quickly after the the voltage of the pre-synaptic cell lowers below the synaptic threshold. A synapse is slow when the decaying current generated by the pre-synaptic cell lasts much longer, so its duration can be compared with the inter-spike/burst period, and not with the spike/burst duration as in the case of fast synapse. This property of long decay is key to understand how two neurons coupled reciprocally with slow inhibitory synapses can suppress in-phase oscillations \cite{van1994inhibition}. Basically, if both are given a ``window of opportunity'' to start together they will continue oscillating in phase. Otherwise, if the initial conditions are different and the neurons do not start within this window, either one may surpass the other if the coupling is strong enough and the decay is long enough. As such, there is no asymptotic convergence to the synchronous state, but to the anti-phase rhythm. This is not the case when one examines HCOs made of endogenous bursters with weak inhibitory coupling, using the fast FTM-synapses, provided that the level of the synaptic threshold goes {\em through} the fast spikes within bursts; if the level is below the spikes, the burster effectively becomes a FN-neuron or a gFN-neuron. It was demonstrated in \cite{jalil2010fast,jalil2012spikes} that in-phase synchrony of two inhibitory coupled HH-like bursters can be stable and asymptotic for three different models of the fast inhibitory synapses. Moreover, such busters can also converge to other close synchrony-like states, with one or several spikes apart, due to the spike timing and interactions, which make inhibition act like excitation; the reader can find further details in the above references.  Increasing the coupling strength breaks down the synchrony arising from weak spike interactions, so the neurons start bursting in alternation.                              

\begin{figure}[ht!]
\begin{center}
\resizebox*{.99\columnwidth}{!}{\includegraphics{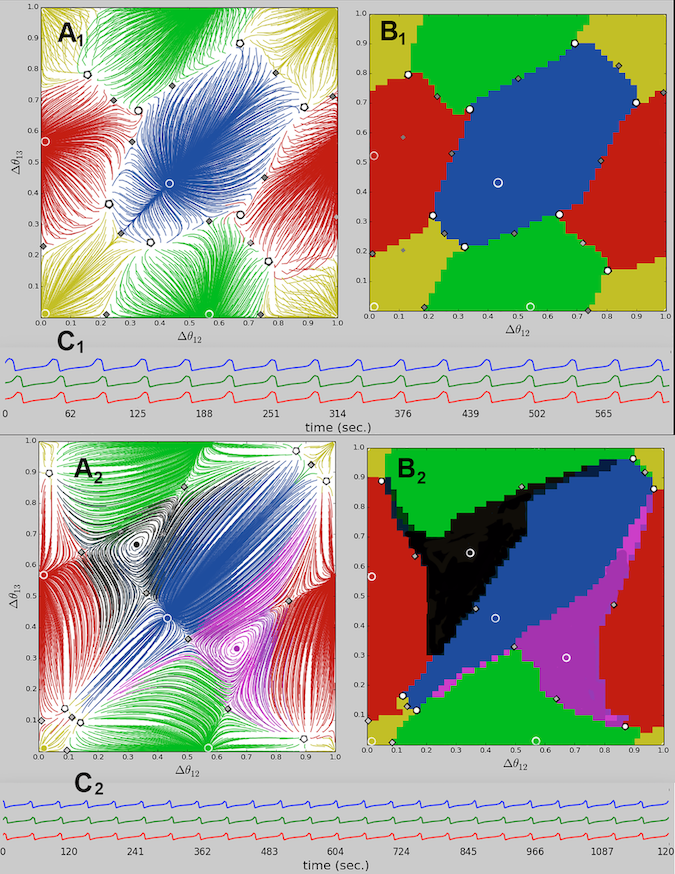}}
\caption{Poincar\'e return maps depicting the stable ``synchronous'' FP at the origin (0,0) (yellow), along with three stable PMs in panel A$_1$ (parameters: $g_{ij}=0.0035$, $I=0.409$, $V_0=-0.113$, $\epsilon=0.3$), or three stable PMs and two stable TWs in panel A$_2$ (parameters: $g_{ij}=0.0021$, $I=0.406$, $V_0=-0.113$, $\epsilon=0.3$). Panels B$_{1,2}$ reveal the attraction basins of the coexisting FPs in greater detail and allow us to determine the locations of the repelling FPs, at the junctions of three distinct color-coded regions. Traces in C$_{1,2}$ demonstrate asymptotic convergence to the synchronous rhythm corresponding to the FP at the origin.
}
\label{fig:sync1}
\end{center}
\end{figure}

	    One can see from Fig.~\ref{fig:sync1} that the FP at the origin in the depicted 2D maps is no longer a repeller, unlike all the previous cases, where it corresponds to an unstable synchronous rhythm with $\Delta_{12}=\Delta{13}=0$. For the given parameters values, the origin becomes an {\em asymptotic} attractor with a relatively large basin (shown in yellow), to which the nearby initial conditions converge. Depending on the gap between the nullclines, the fast $V'=0$ and the slow $h'=0$, this newly formed attractor co-exists with the stable PMs (green, red, and blue) in Fig.~\ref{fig:sync1}A$_1$, or with stable PMs and TWs (black and purple spirals) in Fig.~\ref{fig:sync1}A$_2$. 
\begin{figure}[ht!]
\begin{center}
\resizebox*{.6\columnwidth}{!}{\includegraphics{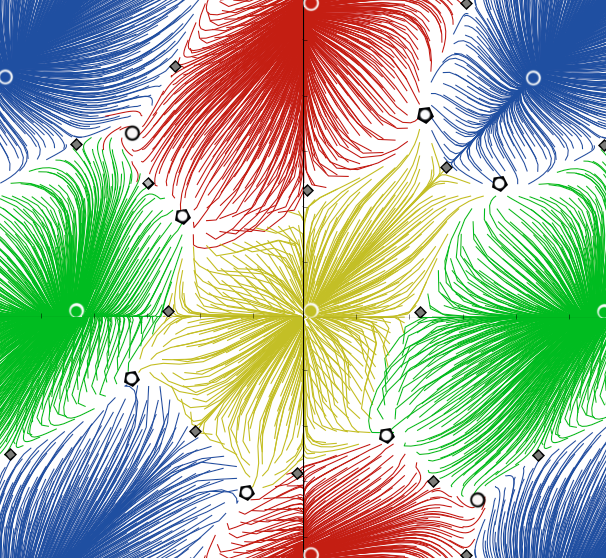}}
\caption{Magnification of the basin of attraction of the stable synchronous state (yellow) in the return map of Fig.~\ref{fig:sync1}A$_1$, by stitching together four identical panels, to disclose its structure with six repellers (white dots) and six saddles (grey diamonds) surrounding the stable FP (yellow)) at the origin (0,0).}
\label{fig:sync2}
\end{center}
\end{figure}
	    This figure also introduces another concept incorporated into MotifToolbox~\cite{motiftoolbox}, to reveal the attraction basins of the co-existing stable FPs (labelled by color-matching dots) in greater detail, as presented in Fig.~\ref{fig:sync1}B$_{1,2}$. Using these diagrams, we can identify the locations of the repellers in the $(\Delta_{12},\Delta_{13})$-plane at the junctions of three distinctively colored regions. For example, in Fig.~\ref{fig:sync1}(A$_1$/B$_1$), one can spot such repellers at the locations of the TW FPs at $\left ( \frac{1}{3},\, \frac{2}{3} \right ) $ and  $\left ( \frac{2}{3},\, \frac{1}{3} \right )$,  as well as six more repellers (white dots) and six new saddles (grey diamonds). The number of repellers in Figs.~\ref{fig:sync1}A$_2$ and B$_2$ provides an explanation as to how the synchronous FP at the origin becomes stable. Due to its location and symmetry on the 2D torus, it undergoes a degenerate pitch-fork bifurcation simultaneously along the lines $\Delta_{12}=0$, $\Delta_{13}=0$ and $\Delta_{12}=\Delta_{13}$, and becomes stable in all three directions. This bifurcation gives rise to six new saddles and six new repellers nearby. Given that the return maps are defined on a phase torus as modulo 1, the particular visualization approach stitching together four identical panels can be employed to magnify the vicinity of the origin, as shown in Fig.~\ref{fig:sync2}. This panel reveals how the boundary of the attraction basin of the synchronous state at the origin are geometrically determined by the stable and the unstable separatrices of the saddles, that emerges through this bifurcation sequence.
\begin{figure}[ht!]
\begin{center}
\resizebox*{.99\columnwidth}{!}{\includegraphics{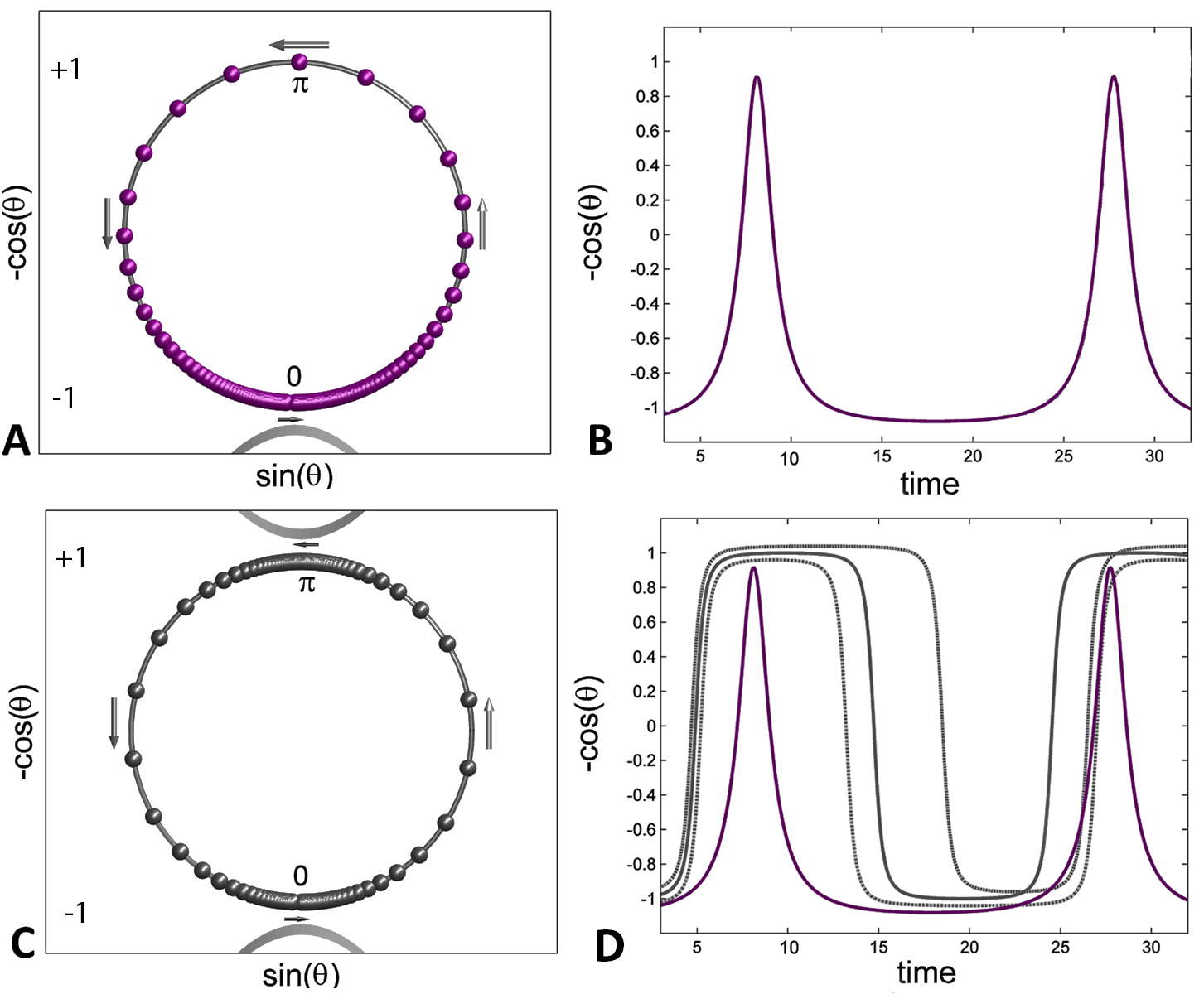}}
\caption{ Unit circle represents the phase space of a $\theta$-neuron with a single ``quiescent'' saddle-node phantom at $0$ (near the quadratic tangency at the bottom) (A) and a $2\theta$-neuron with two single saddle-node phantoms at $0$ (quiescent phase) and $\pi$ (tonic-spiking phase) in (C), where the counter-clockwise rotations slow down as illustrated by the spheres representing the phase points on $\mathbb S^1$. Varying the gaps or the distances from/to the saddle-nodes let one control the inter-spike/burst interval (B) or the duty cycle (D) of bursting traces.}
\label{fig:2theta1}
\end{center}
\end{figure}

\section{Bottom up approach: $2\theta$-bursters for 3-cell motifs}   

The $2\theta$-neuron model is motivated by the dynamics of endogenous bursters, with two characteristic slow phases: tonic-spiking and quiescent. Similar to the notion of the relaxation oscillator vs. the FN-neuron, the equation describing the so-called ``spiking'' $\theta$-neuron \cite{theta} in the context of neuroscience was known for a long time, since it was introduced in the classical mathematical theory of synchronization \cite{synchro}. Its core is a homoclinic Saddle-Node bifurcation on a torus or on an Invariant Circle, or the SNIC bifurcation, that occurs on the V-shaped boundaries of synchronization zones, also known as Arnold tongues, in the parameter plane, see Fig.~\ref{fig:zones}.  The $\theta$-neuron capitalizes on the pivotal property of the saddle-node bifurcation -- the phantom bottle-neck effect that gives rise to slow and fast time scales in the dynamics of systems ranging from simple 1D to higher-order models. In the gFN-model, the saddle-node bifurcation occurs at the quadratic tangency of the nullclines, $V'=0$ and $h'=0$, in the phase plane (Fig.~\ref{fig:keymechs}). Fig~\ref{fig:HH} illustrates the same principle in the 3D phase space of the reduced leech heart interneuron, where the quiescent phase of bursting can be controlled by varying the gap between the slow nullcline and the right hyper-polarized knee. Recall that a similar saddle-node bifurcation in this model, which controls the tonic spiking phase and the number of spikes per burst, is associated with the famous blue-sky catastrophe \cite{Shilnikov2005,blue_schol,shilnikov2012complete,Showcase2014}.

   A key feature of the $2\theta$-neuron is the occurrence of two saddle-node bifurcations, that introduce two slow transitions into its dynamics, with two fast switches in between. Similar to endogenous bursters with two slow transient states -- the active tonic-spiking and the quiescent phases that can be controlled independently, we can manage the durations of the two analogous states in the $2\theta$-neuron: ``on'' at $\pi$  and ``off'' at $0$, using the same bottleneck post-effects of the two saddle-node bifurcations. This allows us to regulate its duty cycle, which is the fraction of the active-state duration compared to the burst period, see Fig.~\ref{fig:2theta1}. As seen in this figure, the $\theta$-model phenomenologically depicts spiking cells, while the $2\theta$-neuron can be treated as a ``spike-less'' burster, similar to the gFN-neuron discussed previously. Below, we demonstrate that the network dynamics produced by a 3-cell motif composed of inhibitory $2\theta$-bursters, preserve all the key features seen in a motif composed of the three gFN-neurons as well.         
      
  \begin{figure}[h!]
\begin{center}
\resizebox*{.99\columnwidth}{!}{\includegraphics{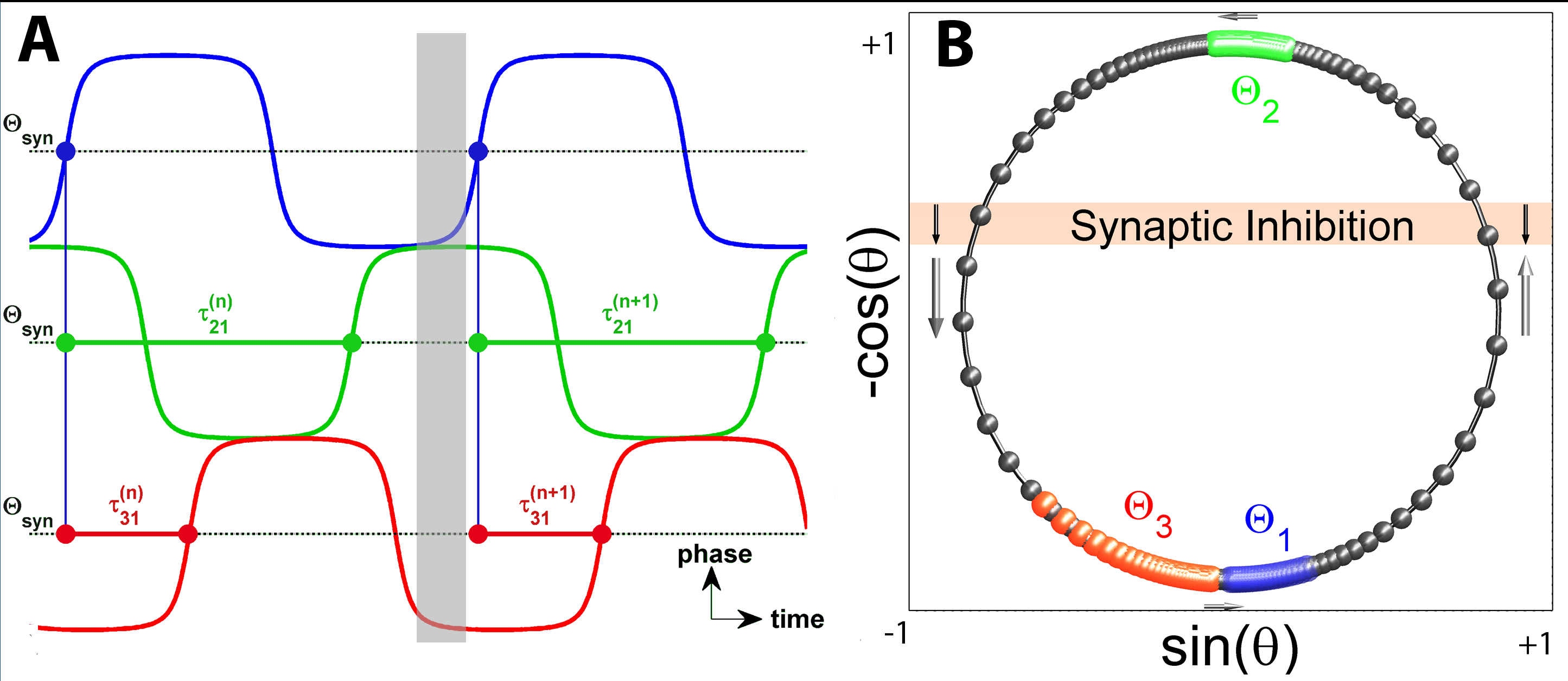}}
\caption{(A) Phase lags between the three $2\theta$-neurons are recorded when the phase/voltage reaches the synaptic threshold $\theta_{\rm syn}=0$ from below. (B) Phase progressions of the coupled $2\theta$-neurons and their color-coded phase points on a unit circle $\mathbb S^1$. }
\label{fig:2theta2}
\end{center}
\end{figure}

A 3-cell motif comprised of $2\theta$-neurons, coupled with fast inhibitory FTM-synapses is given by the following system:

\onecolumngrid
\begin{equation}
\label{eq:2thetaneuron}
\begin{cases}
\theta'_1=\omega-\cos{2\theta_1}+ \textcolor{blue} {\alpha\cos{\theta_1}} -  \displaystyle \left (\frac{\beta_{21}}{1+e^{k\cos{\theta_2}}}+\frac{\beta_{31}}{1+e^{k\cos{\theta_3}}} \right ) \cdot \textcolor{red}{ \left [1-\frac{2}{1+e^{k\sin{\theta_1}}} \right ]},\\~\\
\theta'_2=\omega-\cos{2\theta_2}+ \textcolor{blue} {\alpha\cos{\theta_2}} -  \displaystyle \left (\frac{\beta_{12}}{1+e^{k\cos{\theta_1}}}+\frac{\beta_{32}}{1+e^{k\cos{\theta_3}}} \right ) \cdot \textcolor{red}{ \left [1-\frac{2}{1+e^{k\sin{\theta_2}}} \right ]},\\~\\
\theta'_3=\omega-\cos{2\theta_3}+ \textcolor{blue} {\alpha\cos{\theta_3}} -  \displaystyle \left (\frac{\beta_{13}}{1+e^{k\cos{\theta_1}}}+\frac{\beta_{23}}{1+e^{k\cos{\theta_2}}} \right ) \cdot \textcolor{red}{ \left [1-\frac{2}{1+e^{k\sin{\theta_3}}} \right ]},
\end{cases}
\quad \mbox{mod 1.}
\end{equation}
\twocolumngrid
One can observe that the phase dynamics of an individual $2\theta$-neuron are governed by the terms  $\omega-\cos(2\theta)$. As long as the frequency $0<\omega \le 1$, there are two stable and two unstable equilibria: at the bottom around $\theta \simeq 0$ and at the top near $\theta \simeq \pi$; they are associated with the hyperpolarized and the depolarized quiescent states of the neuron. When $\omega > 1$, the $2\theta$-neuron becomes oscillatory through two simultaneous saddle-node bifurcations (SNIC) on a unit circle $\mathbb  S^1$, where $\theta$ is defined on modulo one. Moreover, whenever $\omega=1+\varepsilon$, where $0<\varepsilon \ll 1$, this new ``burster'' possesses two slow phases: the active ``on'' state near $\theta=\pi$, and the inactive ``off'' state near $0$ on $\mathbb S^1$, alternating with fast counter-clockwise transitions that are referred to as the upstroke and the downstroke, respectively. For greater values of $\omega$, the active and inactive phases should be defined by $\pi/2< \theta \leq 3\pi/2$ and $3\pi/2 < \theta \leq \pi/2$, respectively. The latter phase is below the synaptic threshold which is set by $\theta_{th}=\pi/2$ so that $\cos(\theta_{th})=0$, thus equally dividing the unit circle. The duty cycle of the $2\theta$-neuron is controlled by the term $\textcolor{blue} {\alpha\cos{\theta}}$, provided that it remains oscillatory as long as $\omega-|\alpha|>1$. Note that when $\alpha=0$, the duty cycle is 50\% and the oscillations are even. The active or inactive phases can be extended or shortened, respectively, with $\alpha<0$, making the duty cycle greater, or vice versa -- the duty cycle of individual neurons can be decreased with $\alpha>0$. 

The $2\theta$-neurons are coupled in the network using the fast inhibitory FTM \cite{FTM} synapses. The ``sigmoidal'' term  $\displaystyle \frac{1}{1+e^{k\cos{\theta_i}}}$, ranging between $1$ and $0$, rapidly triggers (here $k=10$) an influx of inhibition flowing from the $i$-th pre-synaptic neuron into the $j$-th post-synaptic neuron, as soon as the former enters the active on-phase above the synaptic threshold $\cos(\theta_{th})=0$, i.e  $\pi/2<\theta_i <3\pi/2$. The strength of the inhibitory coupling is determined by the maximal conductance $\beta_{ij}$ that slows down the rate of increase of $\theta^\prime_{j}$ in the $j$-th post-inhibitory neuron, because of the negative sign of the coupling term. To translate the synaptic input into qualitative inhibition, the sign of the input is switched upon crossing the values $\theta = 0$ and $\theta = \pi$. This is achieved by multiplying all the coupling terms of each ODE by $\textcolor{red}{ \left [1-\frac{2}{1+e^{k\sin{\theta}}} \right ]}$. When $0 <\theta <\pi$, the inhibition is a negative input to slow the transition into bursting. When $\pi <\theta_i <2\pi$, the inhibition is a positive input making the transition out of bursting toward quiescence faster. This is logically consistent as, in general, inhibition should shorten the duration taken for the post-synaptic neuron to leave its active phase. Optionally, one can replace this term with $\left [1-\frac{1}{1+e^{k\sin{\theta}}} \right ]$, that breaks the symmetry was well.

Figure~\ref{fig:2theta2} shows the snapshots of the phase progressions of the three calls on the unit circle $\mathbb S^1$ and depicts how phase lags between the three $2\theta$-neurons are introduced (here, the reference cell is cell 1, in blue), just like in the case of the 3-cell gFN-motif in Fig.~\ref{fig:phaselagreturns}. One can see from Fig.~\ref{fig:2theta2}B that the active green neuron in the active phase near  $\theta=\pi$, above the synaptic threshold, inhibits and pushes the other two closer to each other, near the bottom quiescent state at $\theta=0$, by accelerating the red neuron on the downstroke, and by slowing down the blue neuron on the upstroke.
  \begin{figure}[ht!]
\begin{center}
\resizebox*{.99\columnwidth}{!}{\includegraphics{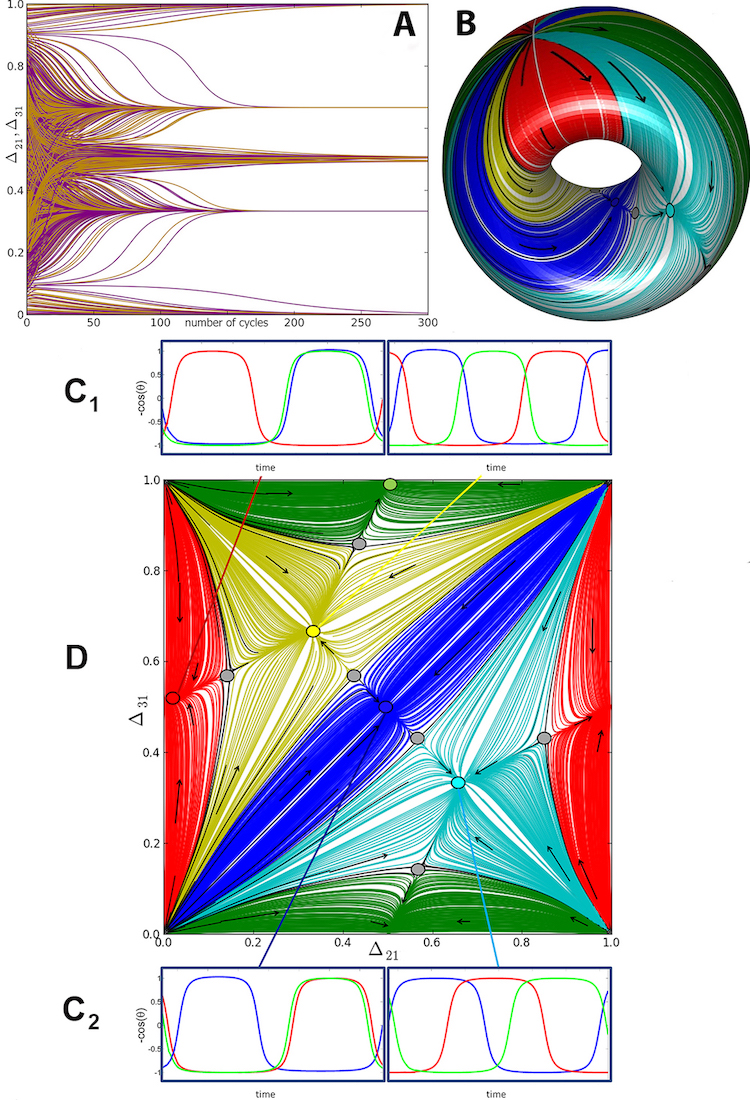}}
\caption{(A) Time progression of multiple initial conditions exponentially converging to several phase-locked states (four of them shown in C1--C2), generated by a symmetric 3-cell motif (see Fig.~\ref{fig:3cellmotif}A). These phase-locked states correspond to the co-existing stable FPs of the 2D Poincar\'e map depicted in panel D, wrapping around the 2D torus in Panel B.}
\label{fig:2theta3}
\end{center}
\end{figure}

Following the same approach used in the weakly coupled HH and gFN models, we use a uniform distribution of initial phase conditions, and hence, the phase lags between the three $2\theta$-neurons and determine the phase locked states that they can converge to, with increasing number of the cycles. This approach is illustrated in Fig.~\ref{fig:2theta3}A (compare with Fig.~\ref{fig:phaselagreturns}) for the symmetric 3-cell motif composed of identical $2\theta$-neurons and equal inhibitory synapses. The corresponding 2D Poincar\'e return map, with the co-existing fixed points and saddles, is shown in Fig.~\ref{fig:2theta3}D, defined on mod~1. By stitching together the opposite sides of this flat map, we wrap it around a 2D torus shown in  Fig.~\ref{fig:2theta3}B, color-coded accordingly. 

One can see that the 2D return maps for the gFN-neurons and $2\theta$-neurons are nearly identical. This implies that our descriptions and modeling approaches to reveal the intrinsic properties of individual neurons and their phenomenological interactions at the network-level, are generic and universal. We would like to underline that the the proposed concept of $2\theta$-bursters bares a great promise for studies of collective dynamics exhibited by larger and modular networks with a combination of inhibitory, excitatory and electric synapses, as well as for modeling biologically plausible circuits such as central pattern generators. To conclude this section, we point out another helpful feature of the $2\theta$-neuron paradigm, namely, the ability to find repelling FPs, if any, in the 2D Poincar\'e map, by reversing the direction of integration of the system~(\ref{eq:2thetaneuron}), i.e., integrating it in the backward time by multiplying the right-hand sides in Eqs.~(\ref{eq:2thetaneuron}) by -1. Unlike the gFN and other HH-like dissipative neural systems where the backward integration will make solutions run to infinity, it is not the case for $2\theta$-bursters, as the phases on $\mathbb S^1$ will just reverse the direction and spin clockwise on the unit circle.

\section{Conclusions and future directions}

The purpose of this paper is to present a de-facto illustration that 3-cell and larger neural networks can universally produce the same emergent behaviors in response to parameter variations, provided that the dynamical properties are properly chosen for the synapses and the constituent neurons, whether those are biologically plausible Hodgkin-Huxley type bursters, reduced generalized Fitzhugh-Nagumo neurons, or toy $2\theta$-bursters. In all these cases, we can employ the reduction to the visually evident Poncar\'e return maps for phase lags, solely derived from multiple voltage traces. This presents a potent computational approach for the thorough analysis of rhythmic behaviors arising in a range of symmetrical and asymmetrical neural networks. By taking advantage of the latest GPU computing paradigms, we perform fast parallel computations of numerous network trajectories and construct the return maps, such as Fig.~\ref{fig:relsymI42} and Fig.\ref{fig:relsymI39}, within just a few seconds. We demonstrate how the reduced 2D gFN neuron model, given by equations (\ref{3cellFHeq}), in conjunction with these computational techniques, allows for comprehensive examinations of rhythmic behaviors arising in these networks, their underlying mechanisms of release and escape, as well as the construction of detailed bifurcation diagrams (Fig.~\ref{fig:relsymgrid}$E$) to study rhythm transitions as the parameters such as the external current drive and the strength of one or more synapses are manipulated. The parameters are carefully chosen so they can also be controlled in neurophysiological experiments with dynamic clamp, to replicate these behaviors in real animal CPGs \cite{sakurai2017artificial, sakurai2014two}. Symmetric and asymmetric 3-cell configurations can produce a range of stable and unstable rhythmic behaviors, including phase-locked bursting with pacemakers or traveling waves, as well as the recurrent phase slipping chimeras. A rich set of bifurcations can be induced in these networks including saddle-node, pitch-fork, and secondary Andronov-Hopf/torus bifurcations through parametric changes, resulting in the emergence/disappearance of rhythmic states, and the gain/loss of their stability. Finally, we also demonstrate how the 3-cell motif can lose all its stable/unstable fixed points and invariant circles, causing the emergence of a transient torus, where the network can produce voltage traces whose phase lags vary in a weakly chaotic manner.  

\begin{figure}[h!]
\begin{center}
\resizebox*{0.8\columnwidth}{!}
{\includegraphics{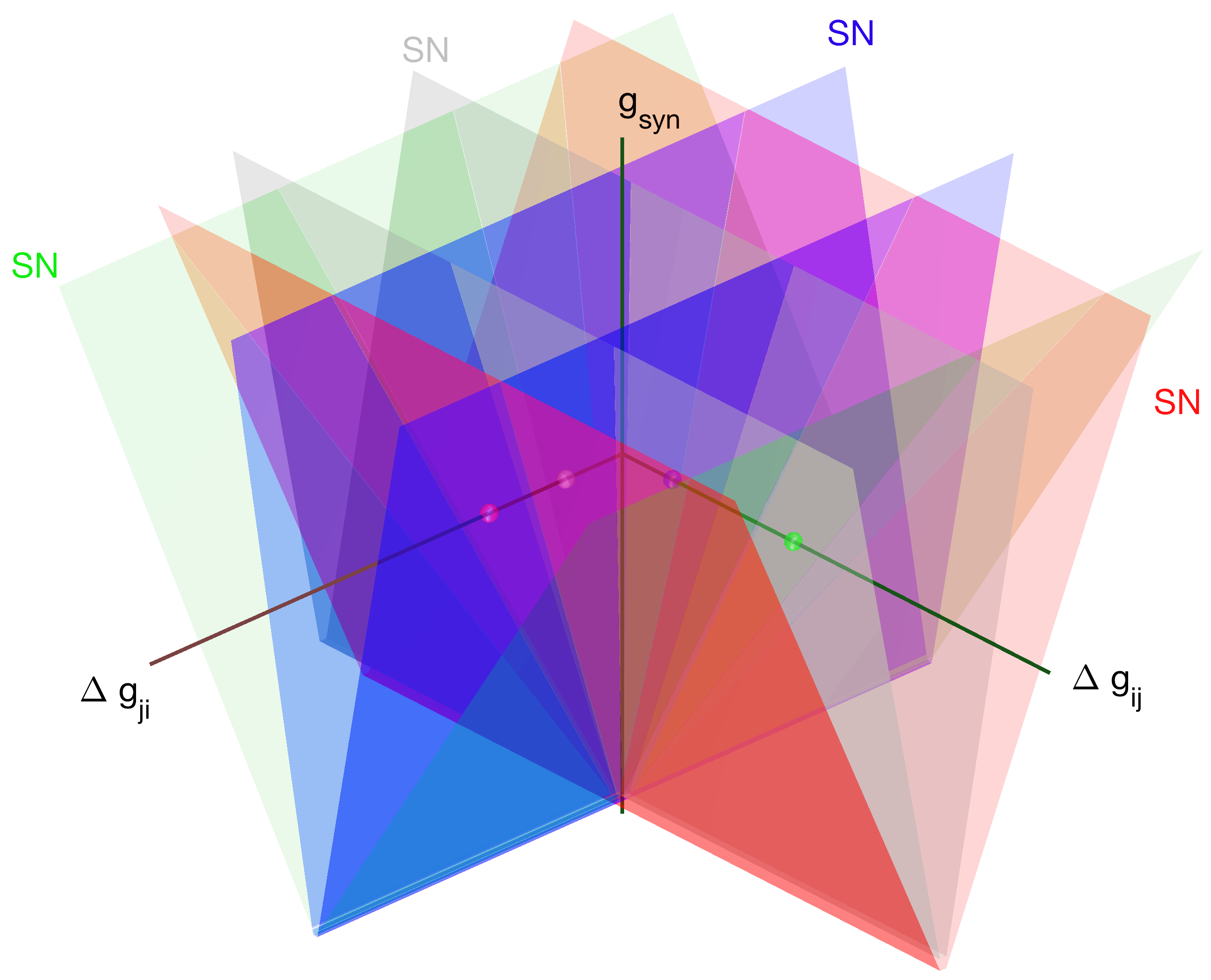}}
\caption{Diagram outlining several nested synchronization zones in the  parameter space of a neural network. Pairs of stable and unstable FPs in the 2D return map are eliminated sequentially when the boundaries of these zones, corresponding to saddle-node (SN) bifurcations, are crossed outwardly as the coupling $\Delta g_{ij}$  between neurons $i$ and $j$ is increased.}
\label{fig:zones}
\end{center}
\end{figure}

We emphasize that 3-cell circuits composed of the 2D gFN-type neurons replicate the multiplicity of rhythmic behaviors and bifurcations seen in more detailed Hodgkin-Huxley type models, \cite{wojcik2014key} albeit at much lower computational costs. The reduction of the analysis of 3-cell network dynamics to 2D phase lag return maps allows for the simple visual inspection of fixed points, invariant circles and their bifurcations. A stable FP representing a phase-locked bursting rhythm, also remains structurally stable under variations of network parameters, as seen in biparametric scans such as Fig.~\ref{fig:bifsingle} and Fig.~\ref{fig:bifdouble}. These scans show the regions of existence and stability for the FPs, also referred to as synchronization zones \cite{synchro}, in the 2D parameter diagrams. The boundaries of such zones correspond to homoclinic and heteroclinic saddle-node bifurcations of fixed points or periodic orbits on an invariant circle; see \citep{ju2018bottom} and the references therein for more details about tori in neural models.  Since there exist several controllable network parameters such as the current drive for each cell, the connection strengths and other dynamical properties of individual synapses, the nested organization of synchronization zones in the higher dimensional parameter space is depicted in Fig.~\ref{fig:zones}. Such zones have also been known as Arnold tongues \cite{boyland1986bifurcations, mcguinness2004arnold} for weakly coupled harmonic oscillators and other complex vibrating systems. Gradually changing a single parameter can cause a cascade of saddle-node bifurcations as the boundaries of the Arnold tongues are crossed, and rhythmic behaviors emerge/disappear as seen in Fig.~\ref{fig:clockrel} or in Fig.~\ref{fig:clockesc}. Although there exist several controllable network parameters, we demonstrate that many of these manipulations produce qualitatively similar dynamics and transitions, highlighting the effectiveness of these reduction tools. 

\begin{figure}[h!]
\begin{center}
\resizebox*{0.9\columnwidth}{!}
{\includegraphics{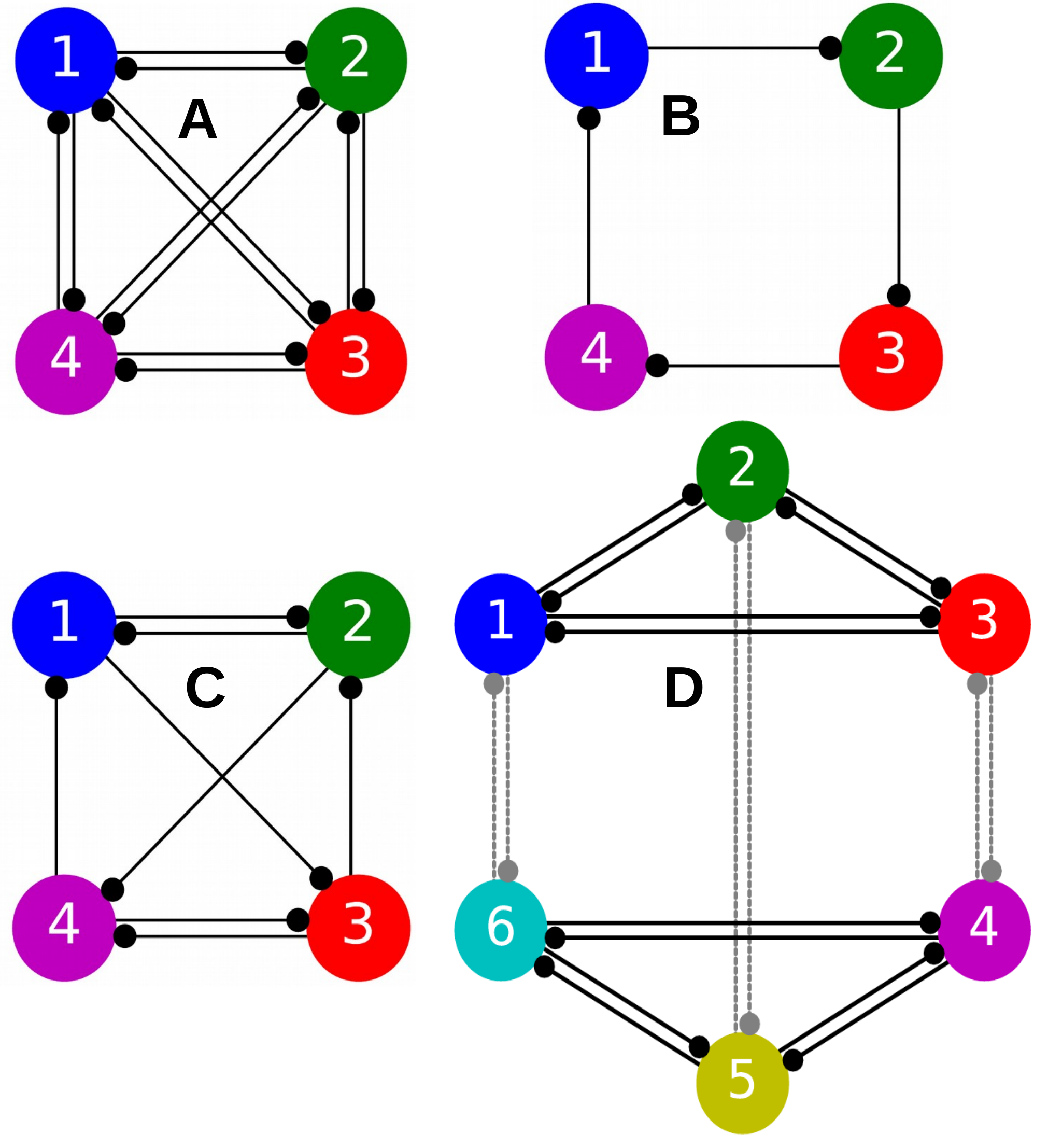}}
\caption{Larger network configurations: The techniques in this study may be extended to the analysis of larger network dynamics, including (A) the multistable fully connected, (B) the bistable one-way inhibitory loop, and (C) the robust monostable mixed 4-cell inhibitory circuits \cite{pusuluri2019computational}, as well as (D) a 6-cell network  composed of two connected 3-cell motifs, reciprocally coupled with cross-inhibitory synapses (dashed gray).
}
\label{fig:larger_networks}
\end{center}
\end{figure}

\begin{figure}[h!]
\begin{center}
\resizebox*{0.9\columnwidth}{!}
{\includegraphics{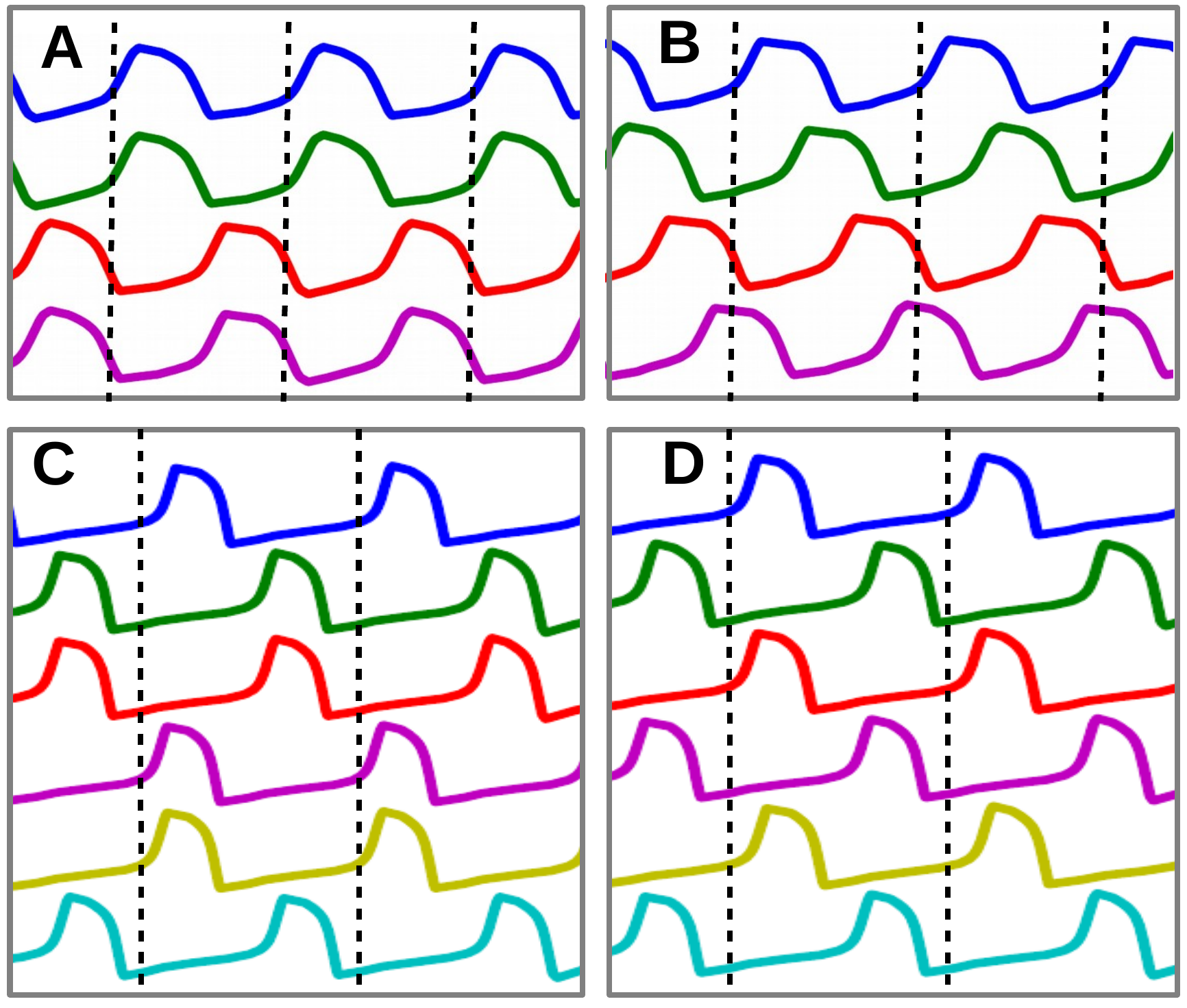}}
\caption{Voltage traces depicting some of the stable rhythms seen in larger inhibitory network configurations:  4-cell networks (see Fig.~\ref{fig:larger_networks}A, B and C) can produce a paired half-center rhythm in (A) or a traveling wave rhythm in (B), among many other outcomes \cite{pusuluri2019computational}; A modular 6-cell network (sketched in Fig.~\ref{fig:larger_networks}D), with 2 coupled 3-cell motifs, cells (1,2,3) and (4,5,6), can generate multiple pacemaker-like rhythms, depicted in Panels (C) and (D), due to the reciprocal cross-inhibition.}
\label{fig:larger_networks_traces}
\end{center}
\end{figure}

The extension of these techniques to study larger neural networks with more than 3 cells as demonstrated in Fig.~\ref{fig:larger_networks} would require additional enhancements, including taking advantage of unsupervised machine learning techniques to analyze the phase lag return maps in higher dimensions. Our preliminary examination of 4-cell inhibitory neural circuits \citep{pusuluri2019computational} and their repertoire of multistable rhythms shows that certain network topologies display a rich array of multistable (Fig.~\ref{fig:larger_networks}A) or bistable (Fig.~\ref{fig:larger_networks}B) rhythms, while others show robust monostability (Fig.~\ref{fig:larger_networks}C), as we vary a range of control parameters. Fig.~\ref{fig:larger_networks_traces}A and B show two such stable rhythms seen in 4-cell inhibitory networks: a paired half-center and a traveling wave. Future work could study the dynamic behaviors arising in larger networks composed of a modular organization of smaller CPG circuits such as the 6-cell circuit shown in Fig.~\ref{fig:larger_networks}D, which is made up of two symmetric 3-cell CPGs that are bound together via reciprocally inhibitory synapses between the corresponding cells in each CPG. The rhythmic outcomes of the larger network depend on the dynamics of both the individual motifs as well as the connections between them. When each 3-cell motif in Fig.~\ref{fig:larger_networks}D is configured to produce just pacemaker rhythms, the cross inhibition (between cells 1--6, 2--5, and 3--4) suppresses certain combinations of pacemaker patterns in the individual motifs, while promoting others as seen in Fig.~\ref{fig:larger_networks_traces}C,~D. Using the known principles and rhythmic outcomes of the smaller CPGs, one might simplify the analysis of such modular networks. Our analysis of phase lags and return maps does not depend on the underlying mathematical equations governing the system. As such, the approach can be generalized to a variety of biological and non-biological complex systems spanning across engineering, economics, population dynamics, dynamic memory and decision making in animals \cite{kee2015feed}, as well as the development of efficient robot locomotion \cite{eckert2015comparing, sprowitz2014kinematic,LShS2019,LShS2019}.

\section*{Acknowledgements}

This study was funded in part by the NSF grant IOS-1455527. We  thank the Brains and Behavior initiative of Georgia State University for the pilot grant support as well as for the fellowships of J. Collens, D. Alacam, and K. Pusuluri.  We thank all the past and current members of the Shilnikov NeurDS (Neuro--Dynamical Systems) lab, particularly J. Scully and J. Bourahman, for inspiring discussions; we also acknowledge S. Kniazev who, using Motiftoolbox\cite{motiftoolbox}, found that the motif could exhibit a stable synchronous state, as a project assignment for {\it Dynamical Foundation of Neuroscience} course offered by AS at GSU. The NeurDS lab is grateful to NVIDIA Corporation for donating the Tesla K40 GPUs that were actively used in this study.

\section*{Data/Code Availability}

The toolkit that supports the findings of this study is openly available as Motiftoolbox \cite{motiftoolbox}.

\section*{References}

\nocite{*}
%

\end{document}